\documentclass{article}
\usepackage{blindtext}
\usepackage{graphicx}

\usepackage{subcaption}
\usepackage[square, numbers]{natbib} 
\usepackage{ragged2e}
\usepackage{amssymb}
\usepackage{graphicx} 
\usepackage{amsfonts}
\usepackage[ruled,vlined]{algorithm2e}
\usepackage{framed}
\usepackage{amsthm}
\usepackage{pdflscape}

\theoremstyle{definition}
\newtheorem{definition}{Definition}

\usepackage{xcolor}

\usepackage[preprint]{neurips_2024}

\graphicspath{{image/}}

\usepackage[utf8]{inputenc} 
\usepackage[T1]{fontenc}    
\usepackage{hyperref}       
\usepackage{url}            
\usepackage{amsmath}        
\usepackage{booktabs}       
\usepackage{amsfonts}       
\usepackage{nicefrac}       
\usepackage{microtype}      
\usepackage{xcolor}         

\title{CosMAP: Contrastive Manifold Approximation and Projection for Dimensionality Reduction of Omics and Genealogical Data}
\author{%
  Fenosoa Randrianjatovo$^{1,2}$ \quad
  Maya Saleh$^{1}$ \quad
  Simon Girard$^{2,3}$ \quad
  Amadou Barry$^{1,2}$\\[0.6em]
  \small
  $^{1}$Institut national de la recherche scientifique (INRS)\\
  Centre Armand-Frappier Santé Biotechnologie (INRS-AFSB),\\
  $^{2}$Unité mixte de recherche en santé durable
  (UMR INRS--UQAC),\\    
  $^{3}$Université du Québec à Chicoutimi (UQAC). \\[1.2em]
  \texttt{\{fenosoa.randrianjatovo, maya.saleh, AmadouDiogo.Barry\}@inrs.ca}\\
  \texttt{simon2\_girard@uqac.ca}
}
\begin{document}

\maketitle

\begin{abstract}
Omics datasets, particularly single-cell RNA sequencing data, are high-dimensional, sparse, noisy, and dominated by zero values, making faithful low-dimensional representation challenging. Existing dimensionality-reduction methods may distort local neighbourhoods, global organization, or the cohesion of meaningful populations, with similar limitations arising in  genealogical data. We introduce Contrastive Manifold Approximation and Projection (CosMAP), a graph-based unsupervised dimensionality-reduction method for producing faithful and interpretable embeddings. CosMAP extends the graph-based framework of UMAP by combining cosine-similarity neighbourhoods with temperature-normalized contrastive affinities, which are optimized in the embedding space using an attractive--repulsive objective. It further employs a two-phase refinement strategy: an intermediate higher-dimensional representation is first learned and then used to reconstruct the neighbourhood graph and initialize the final low-dimensional embedding. We evaluate CosMAP on MNIST and USPS handwritten-digit datasets, mouse retina and cortex single-cell RNA-sequencing datasets, and a large genealogical kinship dataset derived from BALSAC-CARTaGENE. Compared with state-of-the-art dimensionality-reduction methods, CosMAP produces more coherent visual representations, improves neighbourhood preservation, and provides clearer global organization of digit classes, biological cell populations, and regional genealogical patterns. These results indicate that CosMAP offers a robust framework for exploratory analysis of complex, sparse, high-dimensional data. The implementation is publicly available at \url{https://github.com/FenosoaRandrianjatovo/CosMAP-dr}.


\end{abstract}

\section{Introduction}

\noindent Over the past decade, advances in high-throughput sequencing and imaging technologies have transformed how we study complex diseases. In single-cell and multimodal omics, a single experiment can profile tens of thousands to millions of cells, and each cell is represented by expression levels for thousands of genes \cite{jain2016minion}. For example, single-cell RNA sequencing (scRNA-seq) and single-cell ATAC sequencing (scATAC-seq) routinely generate datasets at this scale, so one study can quickly produce matrices that are both massive and high dimensional \cite{qiu2020dropouts,luecken2019current}. These matrices are also intrinsically sparse because many entries are zero due to limited capture efficiency and other technical effects that are often described as dropouts, in addition to genuine low expression \cite{qiu2020dropouts}. In parallel, technologies like the Oxford Nanopore MinION continue to expand genomic data generation by enabling portable and real-time long-read sequencing, with reads reported to exceed 150 kb \cite{jain2016minion}. Across these settings, the bottleneck is no longer data collection but representation. To explore tumour heterogeneity \cite{wang2024tnbc}, compare biological samples across individuals or experimental conditions \cite{dedonno2023population}, and integrate complementary omics modalities \cite{SpatialGlue_2024}, there is a need for unsupervised dimensionality-reduction (DR) methods that transform high-dimensional data into compact embeddings while preserving biologically meaningful structures. Such unsupervised methods are particularly important because reliable labels are often unavailable, incomplete, or defined at a level of granularity that does not necessarily reflect the intrinsic organization of the data. Even when annotations are available, previously unrecognized cell states, tissue domains, or continuous biological transitions may remain hidden within the underlying latent structure. The resulting embeddings should therefore support data-driven biological discovery while remaining computationally efficient and sufficiently interpretable for downstream analysis \cite{luecken2019current}.

\begin{figure*}[!htbp]
\centering
\includegraphics[width=\textwidth]{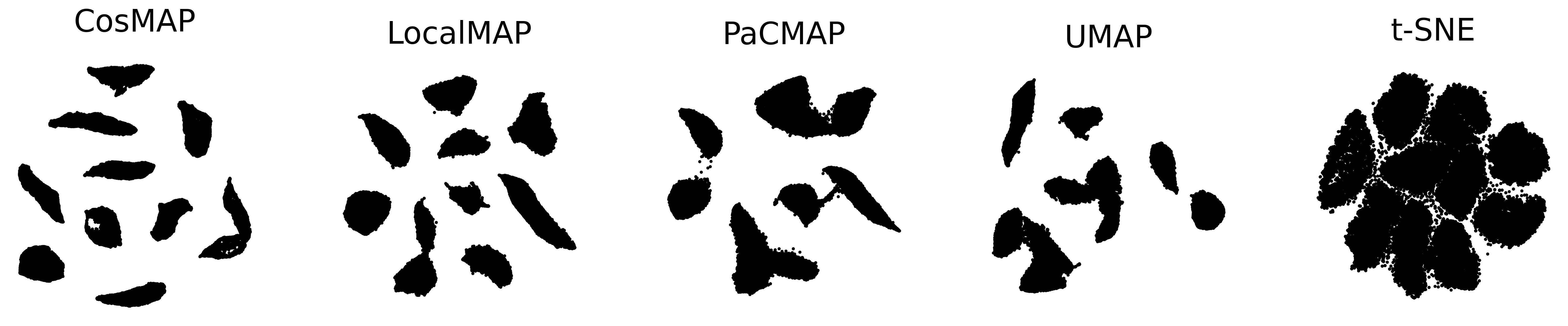}
\caption[Unsupervised visualization of MNIST using different dimensionality reduction methods]{
Unsupervised visualization of the MNIST dataset, used here as a visual motivation for dimensionality reduction. MNIST consists of 70,000 grayscale images of handwritten digits from 0 to 9, each of size $28 \times 28$ pixels. The points are displayed without class coloring to emphasize the latent structure learned by each method. Compared to other methods, the projection obtained with CosMAP highlights a clearer visual separation of the ten-digit populations.
}
\label{fig:mnist_intro_black_intro_paper}
\end{figure*}
\noindent

\noindent
 However, these objectives are difficult to satisfy simultaneously, since improving the separation of some groups may distort their internal organization or alter the broader geometry of the data. In this work, we focus on graph-based dimensionality-reduction methods because they explicitly represent high-dimensional neighbourhood relationships through a graph whose construction and optimization strongly influence the structure visible in the final embedding.

\noindent
Figure~\ref{fig:mnist_intro_black_intro_paper} provides a simple illustration of these challenges. It presents two-dimensional embeddings of MNIST~\cite{Yan_lecun_mnist} produced by CosMAP, LocalMAP, PaCMAP, and t-SNE, with all observations displayed in black. Removing the class colours prevents the known labels from guiding the visual interpretation and makes it possible to examine whether coherent groups emerge from the geometry of the embeddings alone. Although the different methods recover meaningful organization, some representations exhibit overlaps, partial mergers, or ambiguous boundaries. In comparison, the CosMAP embedding reveals a clearer organization of the ten underlying digit populations, despite the fact that no label information is used during optimization. This example motivates the main methodological question considered in this work: How can a neighborhood graph and its low-dimensional optimization be designed within a contrastive-learning framework so that the intrinsic structure of high-dimensional data is represented more clearly and faithfully?


Many state-of-the-art representation-learning systems in machine learning (ML) adopt cosine similarity as the default way to compare high-dimensional vectors. Cosine similarity measures the angle between two vectors and, after $\ell_2$-normalization, reduces to a dot product between unit vectors, which is simple and efficient to compute at scale. Self-supervised contrastive learning frameworks such as SimCLR \cite{simclr} optimize cosine-based similarities between augmented views of the same input to learn robust representations \cite{simclr}, and transformer models, which underpin modern large language models, use scaled dot-product attention \cite{attention2023}. These examples suggest that cosine similarity is a natural and effective choice for comparing high-dimensional, structured vectors, including omics profiles.



\noindent \textbf{Problem statement.} Modern DR methods  typically  (i) derive a $k-$NN graph or affinity matrix using some similarity measure (often Euclidean distance, sometimes cosine or correlation) and (ii) optimize a loss  function, in which the high-dimension similarity  graph is supposed to be fixed during that optimization \cite{wang2025localmap}.
In practice, scientists almost always rely on low-dimensional embeddings from unsupervised DR methods to visualize high-dimensional  data in order to draw some conclusion. However, systematic benchmarks have shown that popular non-linear DR methods such as t-SNE and  UMAP, can substantially distort local and global structure and may be sensitive to hyperparameters, preprocessing choices, and initialization \cite{huang2022evaluation,islam2025shape, watson_how_2022}. A central limitation of many graph-based DR methods lies in the construction of the high-dimensional neighbourhood graph. This graph is   often fixed and treated as the structural scaffold on which the low-dimensional
embedding is optimized. However, in very high-dimensional spaces, this assumption
is fragile. Pairwise  distances may lose contrast as the dimension
increases, a classical manifestation of the curse of dimensionality
\cite{bellman_1959}. Consequently, nearest-neighbor relations inferred from the
raw data can become unstable: some true neighbors may be missed, while some
spurious neighbors may be introduced. This issue is particularly important in
gene-expression and single-cell data, where the choice of dissimilarity or
similarity measure, such as Euclidean distance, correlation, or cosine similarity,
can substantially modify the resulting clustering and neighborhood structure
\cite{jaskowiak2014distances,watson_how_2022}. Recent work has emphasized that the high-dimensional graph should not be regarded
as fully reliable. For example, LocalMAP \cite{wang2025localmap} argues that the initial graph used by
neighbor-embedding methods may contain false-positive edges and proposes to
adjust the graph locally during the optimization in order to reduce the influence
of such unreliable connections \cite{wang2025localmap}. This idea highlights an
important limitation of classical pipelines: the final embedding is only as
reliable as the graph that drives the attractive and repulsive forces. However,
modifying the graph directly during the final two-dimensional optimization may
also affect the geometry of the embedding itself. In particular, recent analyses
of attraction--repulsion dynamics in UMAP and related methods show that changes
in attractive forces can modify cluster formation and may reveal finer structures,
but they can also lead to over-fragmentation when the optimization becomes too
aggressive \cite{islam2025shape}.

\noindent \textbf{Proposed approach and hypotheses.}
Motivated by these limitations, we introduce \emph{Contrastive Manifold Approximation and Projection} (CosMAP), a new unsupervised DR method designed for high-dimensional data. CosMAP is based on the idea that a more reliable embedding can be obtained by combining cosine-based neighborhood construction, contrastive attraction--repulsion principles, and a refinement strategy that reduces the influence of unstable high-dimensional graph edges before the final two-dimensional optimization.

The design of CosMAP is guided by the following hypotheses. First, we hypothesize that high-dimensional data approximately lie on a low-dimensional manifold \cite{monifold_learning_2017}, so that an appropriately constructed neighborhood graph can capture its intrinsic structure. In high-dimensional settings, the choice of proximity metric strongly affects this graph. Cosine similarity is often advantageous because it reduces the influence of differences in vector magnitude and can better capture similarities in expression patterns than Euclidean distance computed directly in the original feature space \cite{watson_how_2022,jaskowiak2014distances}. Consequently, a neighbor-embedding method operating on $k$-NN graphs under cosine similarity is expected to preserve local neighborhoods more faithfully than Euclidean-based $k$-NN graphs in many high-dimensional applications. Second, we hypothesize that contrastive principles can strengthen graph-based dimensionality reduction when they are integrated into both the construction of the affinity graph and the optimization of the embedding. In contrastive representation learning, latent spaces are commonly organized by increasing the agreement between positive pairs and decreasing the agreement between negative pairs \cite{drlim,simclr,attention2023}. Transposed to dimensionality reduction, this suggests that the affinity graph should encode positive neighborhood relations, while the optimization should explicitly introduce repulsive interactions with sampled non-neighboring points. CosMAP follows this hypothesis by coupling a metric-dependent affinity graph with an attraction--repulsion optimization scheme driven by negative sampling \cite{word2vec2013}. Third, although Tessari et al. proposed the Dimension Insensitive Euclidean Metric (DIEM) \cite{tessari2025diem}, which has attractive theoretical properties and can surpass both Euclidean distance and cosine similarity in many high-dimensional regimes \cite{tessari2025diem}, it entails higher computational and implementation costs. We therefore hypothesize that a method based on cosine $k$-NN graphs can offer a favorable balance between embedding quality and computational efficiency for routine omics analyses, so that more complex metrics are not always necessary in practice.

Finally, we hypothesize that high-dimensional neighborhood graphs may contain unstable or spurious edges \cite{wang2025localmap}. Instead of modifying the graph directly during the final two-dimensional optimization, CosMAP introduces a refinement strategy that first learns an intermediate representation and then reconstructs a more reliable graph before optimizing the final low-dimensional embedding. Full methodological details are provided in Section~\ref{methods}.

\section{Related work}

\noindent \textbf{Linear methods.}
Early work on DR in high-dimensional  data largely relied on linear methods such as principal component analysis (PCA) and related factor models. PCA finds orthogonal directions that explain maximal variance in the data \cite{pearson1901_pca} and has been widely used for exploratory analysis, visualization, and denoising of gene expression and single-cell datasets \cite{pearson1901_pca,jolliffe2002pca,jolliffe2016pca}. Classical multidimensional scaling (MDS) similarly seeks a low-dimensional configuration of points whose Euclidean distances approximate a given dissimilarity matrix, often derived from pairwise Euclidean distances in the original space \cite{torgerson1958scaling,kruskal1964mds}. These approaches are computationally efficient and remain a standard preprocessing step in many single-cell workflows, because scientist use  PCA on highly variable genes followed by clustering in the principal component space \cite{townes_feature_2019}. However, linear methods assume that relevant structure can be captured by a linear subspace and operate directly on raw pairwise distances or covariance structure \cite{survey_linear_2014}. As a result, they can struggle when cell states lie on nonlinear manifolds or when local neighborhoods are more informative than global variance, which is typical in heterogeneous tumor and immune datasets \cite{monifold_learning_2017, townes_feature_2019}.

\noindent \textbf{Nonlinear and manifold learning methods.}
To overcome the limitations of linear models, a large class of nonlinear DR and manifold learning methods has been developed. Classical metric MDS can be viewed as a nonlinear method when applied to arbitrary dissimilarities, but it still optimizes a global stress function based on pairwise distances \cite{kruskal1964mds}. Isomap extends this framework by replacing direct Euclidean distances with graph-based geodesic distances: it constructs a neighborhood graph, estimates shortest-path distances along the graph, and then applies MDS to these geodesic distances to recover the manifold’s intrinsic geometry \cite{tenenbaum2000isomap}. Other manifold learning techniques focus more explicitly on local neighborhoods. Locally Linear Embedding (LLE) reconstructs each point as a linear combination of its nearest neighbors, then finds a low-dimensional representation that preserves these reconstruction weights \cite{roweis2000lle}. Laplacian Eigenmaps build an affinity graph and use the spectrum of the graph Laplacian to obtain an embedding that preserves local adjacency structure \cite{belkin2003laplacian}. Diffusion maps model a Markov diffusion process on the data graph and embed points according to diffusion distances, providing a multi-scale notion of similarity that has been applied to high-dimensional biological data, including gene expression and single-cell trajectories \cite{coifman_diffusion_2006}. Conceptually, these methods mark a shift from preserving raw pairwise distances in the original space towards preserving structure encoded in a graph (neighborhood graph or diffusion operator). However, many of them still rely on global eigenvalue  and can be computationally demanding on very large single-cell datasets. \cite{watson_how_2022}

\noindent \textbf{Neighbor embedding methods.}
Neighbor embedding methods take the graph-based perspective further by explicitly constructing probabilistic neighborhoods or affinity graphs and then learning an embedding that best preserves these neighborhoods. Hinton and Roweis introduced Stochastic Neighbor Embedding (SNE) \cite{hinton2002sne}, which, for each point, defines a probability distribution over neighbors in the high-dimensional space and seeks a low-dimensional configuration whose neighbor probabilities match the originals as closely as possible \cite{hinton2002sne}. However, because SNE uses Gaussian kernels in both spaces, nearby points in the high-dimensional space tend to be mapped into a much smaller low-dimensional volume, so many points ``crowd'' together and it becomes difficult to represent all local neighborhoods faithfully, this issue  is known as the crowding problem \cite{vandermaaten2008tsne}. t-distributed Stochastic Neighbor Embedding (t-SNE) addresses this by using a heavy-tailed Student $t$-distribution in the low-dimensional space, which allocates more area for moderately distant points and thus alleviates crowding; as a result, t-SNE has become a widely adopted visualization tool for single-cell and other omics data \cite{vandermaaten2008tsne}. Subsequent work has focused on scalability and the trade-off between local and global structure. LargeVis constructs an approximate $k$-nearest neighbor graph and then optimizes a probabilistic objective to place graph nodes in low dimensions, enabling neighbor embeddings for millions of points \cite{tang2016largevis}. TriMap uses triplet constraints (``point $i$ should be closer to $j$ than to $k$'') to better preserve global relationships between clusters while remaining scalable \cite{amid_trimap_2022}. UMAP (Uniform Manifold Approximation and Projection)  \cite{umap_2020} combines manifold-learning ideas with neighbor embedding: in the high-dimensional space, it constructs a fuzzy simplicial set which is a weighted $k$-nearest neighbor graph, and then optimizes a cross-entropy objective so that a corresponding fuzzy graph in the low-dimensional space matches the original as closely as possible. Neg-t-SNE \cite{damrich2022t} uses negative sampling strategy used in UMAP during the optimization of KL divergence between the low and high dimesional similarity.  PaCMAP (Pairwise Controlled Manifold Approximation) \cite{wang2021pacmap} analyzes the design choices in neighbor-embedding losses and proposes a pair-based objective with different categories of pairs (neighbors, mid-near, and far) to jointly preserve local neighborhoods and broader global arrangements \cite{wang2021pacmap}. More recently, Wang et al.\ introduced LocalMAP, a related DR method that dynamically and locally adjusts the $k$-NN graph by extracting more reliable subgraphs and updating edges on the fly, allowing it to separate clusters that other methods such as t-SNE, UMAP, TriMAP, and PaCMAP may merge \cite{wang2025localmap}. However,  in 2026, Islam et al. \cite{islam2025shape} has done an independent  sensitivity analysis of  LocalMAP. They showed that when the neighborhood size is set too small (for example, $k=\text{$5$ or $7$}$ while the default value is $10$), attractive forces can dominate and previously coherent clusters may fragment into multiple pieces in the embedding, indicating that LocalMAP remains sensitive to hyperparameter choices \cite{islam2025shape}.

 \noindent \textbf{Deep generative embeddings.}
In parallel, deep generative models provide a different route to DR, in which the low-dimensional representation is the latent space of a probabilistic model. Variational autoencoders (VAEs) learn a nonlinear encoder--decoder pair together with a latent variable model by maximizing a variational lower bound on the data likelihood \cite{kingma2013vae}. In single-cell RNA-seq analysis, methods such as single-cell variational inference (scVI) \cite{lopez2018deep} and related VAE-based models extend this framework to handle UMI counts and batch effects, using the learned latent space as a low-dimensional representation of cells \cite{gayoso_scvi_tools_2021}. Similarity-assisted VAE (saVAE) \cite{savae_2023} explicitly incorporates graph-based similarity information by adding a pull--push regularization term based on the UMAP loss into the VAE objective, thereby combining model-based and similarity-based DR in a single latent space \cite{savae_2023}. Related approaches such as VAE-SNE integrate the Kullback--Leibler divergence from t-SNE into the VAE loss, encouraging latent representations that better reflect t-SNE-style neighborhood structure \cite{vae_sne}. Multi-modal extensions such as scMVP jointly embed paired single-cell modalities (e.g.\ RNA and chromatin accessibility) in a shared latent space, providing generative representations of complex omics experiments \cite{li2022deep}. In practice, these models often use latent spaces of dimension greater than two for generative modeling and then apply a secondary DR method \footnote{2D embedding from UMAP, PacMAP, LocalMAP or t-SNE} to visualize the 2D latent space \cite{dean2021pepvae, savae_2023}, that suggests that those deep generative methods are very expensive in terms of space and time complexity. Overall, these deep-learning approaches differ from neighbor embedding and classical manifold learning in that they prioritize probabilistic generative modeling and uncertainty quantification; the embedding arises as a latent variable rather than being directly optimized to match a specific similarity graph. In this work, we therefore focus on graph-based neighbor-embedding methods and do not benchmark CosMAP against deep generative embedding models.

\noindent \textbf{Contrastive neighbor embedding.} More recently, several works have recognized that many neighbor-embedding methods can be interpreted through the lens of contrastive learning, where embeddings are trained to pull “positive’’ pairs (neighbors) together and push “negative’’ pairs (non-neighbors) apart. Damrich and Hamprecht \cite{damrich2022t} show that the losses of t-SNE, UMAP, LargeVis, Neg-t-SNE and related methods all lie on a common attraction–repulsion spectrum and can be written as contrastive objectives \cite{damrich2022t}. In deep representation learning, frameworks such as SimCLR \cite{simclr} optimize a normalized temperature-scaled cross-entropy loss (NT-Xent)\footnote{ See the equation (\ref{eq:ntxent-simclr}) in methodological section.}over embeddings produced by an encoder and a small projection head, yielding powerful low-dimensional representations that are mainly used as features for downstream tasks rather than as direct visualizations \cite{simclr}. Building on this viewpoint, contrastive objectives have been designed explicitly for dimensionality reduction: Contrastive Learning with Similarity Enhancement for Dimensionality Reduction (CLSDR) \cite{yang2025clsdr} combines neighborhood-preserving terms with an InfoNCE-style contrastive loss to improve the quality of 2D embeddings \cite{yang2025clsdr}, while  Noise Contrastive Approach for Scalable Visualization (NCVis) \cite{artemenkov2022ncvis} adapt contrastive losses to learn the high-dimensional node embeddings \cite{bohm2025nodeemb, artemenkov2022ncvis}. These developments emphasize the close connection between neighborhood-preserving DR and contrastive learning.

\noindent \textbf{Contributions.}
This work makes the following contributions. First, we introduce \emph{Contrastive Manifold Approximation and Projection} (CosMAP), a new unsupervised dimensionality-reduction method designed for high-dimensional  data. CosMAP constructs affinity graphs\footnote{By default, we use cosine similarity to build the k-NN graph} from  high-dimensional space. These affinities are transformed into a temperature-controlled probabilistic graph, which regulates the strength and sharpness of local neighborhood relationships. The embedding is then learned by minimizing a binary cross-entropy objective between the high-dimensional affinity graph and a low-dimensional heavy-tailed kernel. Second, we propose a two-stage refinement strategy, in which an intermediate higher-dimensional embedding is first learned and then used to construct a more reliable graph for the final two-dimensional representation. Third, we evaluate CosMAP on multiple high-dimensional datasets and compare it with established DR methods in order to assess its ability to preserve both local and global  neighbourhoods, reveal meaningful cluster structure, and reduce sensitivity to unreliable high-dimensional graph construction.


\noindent Among these contributions, the refinement strategy plays a central role in addressing one of the main limitations of graph-based dimensionality-reduction methods: the fact that the neighborhood graph constructed from the original high-dimensional space may be noisy, unstable, or partially unreliable \cite{wang2025localmap}. Rather than assuming that this initial graph perfectly represents the intrinsic structure of the data, CosMAP treats it as an approximate structural prior that can be improved before producing the final visualization. CosMAP addresses this problem through a refinement strategy that treats the
initial high-dimensional graph as a noisy estimate rather than as a ground-truth
structure. Let \(G_X=(V,E_X,P^{(0)})\) denote the first similarity graph built from
the original data matrix \(X\). Instead of directly relying on this graph to
produce the final two-dimensional embedding, CosMAP first learns an intermediate
representation
\[
    X \longmapsto Y_r \in \mathbb{R}^{n\times r},
    \qquad 2 < r \ll D,
\]
where \(D\) is the original  dimension and \(r\) is a moderate
intermediate dimension. This first stage acts as a denoising step: it preserves
the dominant neighborhood information encoded by the initial graph while reducing
the effect of unstable high-dimensional distances and isolated false-positive
edges. A second graph \(G_{Z_r}=(V,E_{Z_r},P^{(r)})\) is then reconstructed from
the intermediate representation \(Y_r\). The final two-dimensional embedding is optimized from this refined graph: 

\[
    X
    \xrightarrow[\text{first CosMAP stage}]{d=r}
    Y_r
    \xrightarrow[\text{rebuild } k\text{-NN graph}]{}
    G_{Y_r}
    \xrightarrow[\text{final CosMAP stage}]{d=2}
    Y \in \mathbb{R}^{n\times 2}.
\]

This design moves graph correction upstream, before the final 
stage. In contrast to methods that repeatedly alter the graph directly inside  the optimization, CosMAP first obtains a more stable intermediate
geometry and then rebuilds the neighborhood graph from this denoised
representation. The final stage therefore does not need to relearn the entire
structure from scratch; it only refines the embedding using a graph whose
neighborhood relations are expected to contain fewer spurious edges. In practice,
this second stage can be performed with a smaller number of epochs, since the
intermediate representation already provides a structured initialization and a
more reliable graph. Thus, CosMAP reduces the dependence on an untrusted
high-dimensional graph while avoiding excessive local graph manipulation in the
final two-dimensional space.

Finally, beyond the methodological contribution, CosMAP is released as an
open-source Python package following the \texttt{scikit-learn} estimator
interface \cite{pedregosaScikitlear_2018}. This makes the method accessible to the broader machine-learning
community through standard \texttt{fit} and \texttt{fit\_transform} workflows,
facilitates its integration into existing preprocessing, benchmarking, and
visualization pipelines, and supports transparent and reproducible comparison
with other unsupervised DR methods.

\section{Methodology} \label{methods}
Assume that  we have $\boldsymbol{X} = \{x_i\}_{i=1}^{n} \subset \mathbb{R}^D$ high dimensional data points that we treated as random variable drawn from an unknown distribution, where $D$ is the original number of dimensions and $d<D$ is the target lower dimension, while preserving as much as possible the intrinsic structure of the data. When \(D\) is large, standard distance-based notions of proximity become less discriminative, a well-known manifestation of the curse of dimensionality \cite{bellman_1959, aggarwal2001surprising}. Thus, rather than attempting to preserve all high-dimensional pairwise distances, CosMAP uses a Temperature-scaled normalization-based (NT-Xent) \ref{eq:ntxent-simclr} affinities to construct a sparse  similarity graph under a given metric, which is then optimized in the low-dimensional embedding.



\subsection{High-dimensional similarity}
\label{sec:method}

CosMAP high-dimensional similarity  is inspired by the temperature-scaled normalization (NT-Xent) used in  self-supervised  contrastive learning, but it does not adopt the full image-based contrastive learning framework. In methods such as SimCLR \cite{simclr}, the contrastive task is constructed from data augmentations: two augmented views of the same image form a positive pair, while other images in the minibatch act as negatives \cite{simclr}. The corresponding NT-Xent loss encourages representations of the two views to be close and representations of different images to be separated:
\begin{equation}
        \ell_{i,j}
    = - \log
    \frac{\exp\bigl(\mathrm{sim}(z_i, z_j)/\tau\bigr)}
         {\sum_{k \in \mathcal{B}\setminus\{i\}}
          \exp\bigl(\mathrm{sim}(z_i, z_k)/\tau\bigr)} ,
    \label{eq:ntxent-simclr}
\end{equation}
where \(z_i\) and \(z_j\) are projected neural representations, \(\mathcal{B}\) is the minibatch, \(\mathrm{sim}(\cdot,\cdot)\) is  cosine similarity, and \(\tau>0\) is a temperature parameter.

Such contrastive methods were originally designed for representation learning, especially in computer vision, where a nearest-neighbour graph built directly in pixel space is often not semantically meaningful. For example, two images may be close in pixel space because of background, brightness, or low-level texture, while belonging to different semantic classes. This limitation motivates methods such as t-SimCNE \cite{tsimcne_2023}, which combine contrastive learning and neighbour embedding to reduce and to visualize image datasets by learning a parametric mapping into two dimensions \cite{tsimcne_2023}. In that setting, semantic neighbourhoods are obtained through augmentation-based contrastive learning rather than through a direct k-nearest-neighbour graph in raw pixel space.

CosMAP differs from these image-oriented contrastive methods. It does not use image augmentations, a pretrained image encoder, or a neural projection head to define positive pairs. Instead, positive pairs are defined directly by the \(k\)-nearest-neighbour structure of the data under some metrics of similarity; by default we use cosine similarity. This choice is motivated by the structure of sparse omics data, where the observations already live in a biologically meaningful feature space, and where angular similarity between normalized expression profiles provides a natural notion of cell--cell similarity. Although, the default metric used to compute the  k-NN graph in high dimesional is cosine similarity, cosine and Euclidean distance are closely related once vectors are $\ell_2$-normalized. For two unit vectors $x_i$ and $x_j$,
\begin{equation}
    \|x_i - x_j\|^2
    = \|x_i\|^2 + \|x_j\|^2 - 2 x_i^\top x_j
    = 2 - 2 \mathrm{sim}(x_i,x_j),
\end{equation}
so that
\begin{equation}\label{eq:cos_vs_euc}
    \mathrm{sim}(x_i,x_j)
    = 1 - \frac{\|x_i - x_j\|^2}{2}.
\end{equation}

\noindent For each observation \(x_i\), if the metric is cosine, then we first normalize so that \(\mathrm{sim}(x_i,x_j)=x_i^\top x_j\) is the cosine similarity between \(x_i\) and \(x_j\). Let \(\mathcal{N}_k(i)\) denote the set of \(k\) nearest neighbours of \(x_i\) under that  given   metric. CosMAP then defines a local, temperature-scaled neighbourhood distribution by
\begin{equation}
\label{eq:pij}
    P_{j\mid i} =
    \begin{cases}
        \displaystyle
        \frac{\exp\!\bigl(\mathrm{sim}(x_i, x_j)/\tau\bigr)}
             {\displaystyle\sum_{l\in\mathcal{N}_k(i)}
             \exp\!\bigl(\mathrm{sim}(x_i, x_l)/\tau\bigr)},
        & j \in \mathcal{N}_k(i),\\[0.8em]
        0, & \text{otherwise}.
    \end{cases}
\end{equation}
Here, \(\tau>0\) controls the sharpness of the local similarity distribution. Smaller values of \(\tau\) concentrate the probability mass on the most similar neighbours, whereas larger values produce a smoother distribution over the \(k\)-neighbourhood.

Thus, CosMAP keeps the temperature-scaled normalization principle from self-supervised contrastive learning above, but replaces augmentation-defined positives by graph-defined positives. Instead of contrasting an anchor against all other samples in a minibatch, CosMAP normalizes only over the  neighbourhood \(\mathcal{N}_k(i)\). This makes the construction closer to graph-based neighbour embedding methods such as NCVis \cite{artemenkov2022ncvis}, while retaining a contrastive normalization mechanism through the temperature parameter.

Finally, since \(P_{j\mid i}\) is directional, we define the symmetric high-dimensional edge strength by
\begin{equation}
\label{eq:pij_sym}
        p_{ij}
    =
    \frac{P_{j\mid i}+P_{i\mid j}}{2},
    \qquad
    p_{ii}=0.
\end{equation}
Setting the diagonal elements $p_{ii}$ to zero optimizes computational efficiency, forcing the algorithm to exclusively evaluate the interactions between distinct pairs of points. The resulting matrix \(P = (p_{ij})\) is sparse and symmetric, and can be interpreted as a weighted adjacency matrix of size \(n \times n\).

\subsection{Low-dimensional similarity}

In the low-dimensional embedding, we  model pairwise similarities with a heavy-tailed kernel to approximately compensate for the crowding problem \cite{umap_2020}. Intuitively, a heavy-tailed distribution allocates more “space’’ for moderately distant points, preventing all neighbours from being compressed into a tiny region around each point, as happens when using Gaussian kernels in low dimensions \cite{vandermaaten2008tsne,umap_2020}. 
From equation \ref{eq:cos_vs_euc}, the same heavy-tailed functional form that UMAP uses for Euclidean distances in the embedding space is naturally compatible with our approach in the input space: after normalization, the cosine kernel corresponds to a simple quadratic transform of the Euclidean distance.

This allows CosMAP to use a UMAP-style heavy-tailed kernel for low-dimensional similarities while keeping an interpretation of neighbourhood structure in the original  space.

\noindent Hence, given an embedding $\boldsymbol{Y} = \{y_i\}_{i=1}^n \subset \mathbb{R}^d$, we define a symmetric low-dimensional similarity
\begin{equation}
\label{eq:qij}
    q_{ij} = \frac{1}{1 + a \,\|y_i - y_j\|^{2b}},
    \qquad q_{ii} = 0,
\end{equation}
where $\|\cdot\|$ denotes the Euclidean norm, and $a>0$, $b>0$ are shape parameters controlling the tail behaviour of the kernel. This formulation places CosMAP within the family of neighborhood-preserving dimensionality reduction methods based on attraction--repulsion mechanisms. In particular, the low-dimensional similarities $q_{ij}$ play a role analogous to the probabilistic affinities used in t-SNE, where nearby points are encouraged to remain close while dissimilar points are pushed apart \cite{vandermaaten2008tsne}. At the same time, CosMAP follows the graph-based philosophy of UMAP \cite{umap_2020} by relying on a $k$-nearest-neighbor structure to define the relevant high-dimensional relationships.

\subsection{Optimization}

Given the high-dimensional similarities \(p_{ij}\) and the low-dimensional similarities \(q_{ij}\), CosMAP learns the embedding \(\boldsymbol{Y}=\{y_i\}_{i=1}^{n}\subset \mathbb{R}^{d}\)
by minimizing the following binary cross-entropy (BCE) objective:
\begin{equation}
\label{eq:cross_entropy}
    \mathcal{L}
    =
    -\sum_{i \neq j}
    \Bigl[
        p_{ij}\log q_{ij}
        +
        (1-p_{ij})\log(1-q_{ij})
    \Bigr].
\end{equation}

This objective can be interpreted through the standard attraction--repulsion perspective used in graph-based neighbour-embedding methods. This interpretation is not specific to CosMAP: related attraction--repulsion mechanisms appear in stochastic neighbour embedding and t-SNE, where similar points are attracted while dissimilar points are repelled in the embedding space~\cite{hinton2002sne, vandermaaten2008tsne}; in LargeVis, where a graph-layout objective is optimized using negative sampling~\cite{tang2016largevis}; and in UMAP, where a fuzzy-set cross-entropy objective is optimized by attractive updates on observed edges and repulsive updates on sampled negative edges~\cite{umap_2020}. In CosMAP, this standard mechanism is instantiated using the high-dimensional affinity \(p_{ij}\) defined in equation~\eqref{eq:pij} and the low-dimensional edge strength \(q_{ij}\).

Indeed, the BCE objective in equation~\eqref{eq:cross_entropy} contains two complementary contributions:
\begin{equation}
    \mathcal{L}= \mathcal{L}_{\mathrm{attr}} +\mathcal{L}_{\mathrm{rep}},
\end{equation}
where
\begin{equation}
    \mathcal{L}_{\mathrm{attr}}= -\sum_{i\neq j} p_{ij}\log q_{ij}, \qquad \mathcal{L}_{\mathrm{rep}} = -\sum_{i\neq j} (1-p_{ij})\log(1-q_{ij}).
\end{equation}
The first term is dominant for pairs with large high-dimensional affinity \(p_{ij}\). For such pairs, reducing the loss requires increasing \(q_{ij}\), which encourages strongly connected observations in the high-dimensional graph to remain close in the embedding. This gives the attractive component of the objective. Conversely, the second term is dominant for pairs with small \(p_{ij}\). For these pairs, reducing the loss requires decreasing \(q_{ij}\), thereby penalizing embeddings in which weakly connected or unrelated observations are placed too close together. This gives the repulsive component of the objective.

The quantities \(p_{ij}\) and \(q_{ij}\) can  be interpreted as edge strengths. In particular, as in UMAP, the equation~\eqref{eq:cross_entropy} is best understood as an edge-wise binary cross-entropy between high-dimensional and low-dimensional graph relations, rather than as a Kullback--Leibler divergence between two normalized probability distributions.

In principle, the repulsive contribution involves all pairs that are weakly connected or absent from the positive neighbourhood graph. Evaluating this full repulsive sum would require considering \(O(n^2)\) pairs, which is computationally prohibitive for large  datasets. CosMAP therefore adopts negative sampling as a stochastic approximation to the full repulsive term. This strategy follows the general idea popularized in word2vec \cite{word2vec2013}, where negative sampling was introduced as an efficient alternative to evaluating all possible negative examples~\cite{word2vec2013}, and later used in large-scale graph-layout and manifold-learning methods such as LargeVis and UMAP~\cite{tang2016largevis, umap_2020}.

More precisely, during stochastic optimization, CosMAP applies attractive updates to positive graph edges and repulsive updates to a small number of sampled negative pairs. The negative pairs are sampled from observations that are not treated as positive neighbours of the current point. Hence, instead of explicitly evaluating the repulsive term over all non-neighbouring pairs, the algorithm estimates its effect through a small set of sampled negatives \cite{umap_2020, word2vec2013}. This preserves the qualitative attraction--repulsion behaviour of the full BCE objective while making the optimization scalable. 
Before the stochastic BCE optimization, CosMAP can be initialized using a spectral embedding of the symmetrized affinity graph \(P=(p_{ij})\). This initialization provides a classical graph-based starting point that places strongly connected vertices close to one another in the initial low-dimensional space. Spectral embeddings are widely used in spectral clustering and manifold learning, where low-dimensional coordinates are obtained from eigenvectors of a graph Laplacian, typically associated with the smallest non-trivial eigenvalues \cite{belkin_niyogi_2001}. In a similar spirit, UMAP also uses a spectral initialization of its neighborhood graph before optimizing its low-dimensional objective~\cite{umap_2020}. In CosMAP, this initialization can also be combined with the proposed refinement strategy, where an intermediate embedding is used to improve the graph structure before the final low-dimensional optimization. 

\begin{definition}[Two-phase refinement strategy  in CosMAP]

Let $\boldsymbol{X}=\{x_i\}_{i=1}^{n}\subset\mathbb{R}^D$ be the input data, and
let \(d<r<D\). Let \(\operatorname{CosMAP}_{m}(Z;Y_0)\) denote the output of the
\(m\)-dimensional CosMAP optimization applied to the data \(Z\), initialized at
\(Y_0\). When the initialization is not specified, we write
\(\operatorname{CosMAP}_{m}(Z)\) and use the default spectral initialization.

\noindent The two-phase CosMAP procedure is defined as follows. First, an intermediate
\(r\)-dimensional representation is learned from the original data:
\begin{equation}
\label{eq:intermediate_cosmap}
    Y^{(r)}=\operatorname{CosMAP}_{r}(\boldsymbol{X}).
\end{equation}
Second, this intermediate representation is used as the input of a new CosMAP
optimization in dimension \(d\). The second phase is initialized by the
coordinate projection
\begin{equation}
\label{eq:projected_initialization}
    Y^{(d)}_0=\Pi_dY^{(r)},
\end{equation}
where \(\Pi_d:\mathbb{R}^r\to\mathbb{R}^d\) denotes the projection onto the first
\(d\) coordinates. The final embedding is then defined by
\begin{equation}
\label{eq:two_phase_cosmap}
    Y^{(d)}
    =
    \operatorname{CosMAP}_{d}
    \left(
        Y^{(r)};\,Y^{(d)}_0
    \right).
\end{equation}

\noindent Thus, both the first and second phases are nonlinear CosMAP projection
procedures. The coordinate projection \(\Pi_dY^{(r)}\) is only used to initialize
the second phase; it is not itself the final embedding. 
\end{definition}
In practice, the intermediate dimension $r$ is chosen to be larger than the final embedding dimension $d$, while fewer epochs are allocated to the second optimization phase. In our implementation, we set $r=30$ and $d=2$ for planar visualization. The second phase is run for only one quarter of the number of epochs used in the first phase, since most of the structural information has already been captured in the intermediate representation.

\section{Results and Discussions}

In this section, we evaluate the empirical behaviour of CosMAP on several datasets  with different structures and levels of complexity.  Given the large number of dimensionality reduction methods proposed in the literature, a comprehensive comparison with all existing approaches is beyond the scope of this work. Instead, we adopt a targeted evaluation strategy by comparing CosMAP with methods that share a similar objective: producing low-dimensional embeddings for visualization while preserving neighborhood relationships and revealing meaningful structures in high-dimensional data. Accordingly, we focus on widely used and recent nonlinear dimensionality reduction methods designed for data visualization, including t-SNE~\cite{vandermaaten2008tsne}, UMAP~\cite{umap_2020}, PaCMAP~\cite{wang2021pacmap}, LocalMAP~\cite{wang2025localmap}, PHATE~\cite{moon2019phate}, TriMAP~\cite{amid_trimap_2022},  NCVis~\cite{artemenkov2022ncvis}, h-NNE~\cite{sarfraz2022hnne}, and contrastive variants of t-SNE such as InfoNCE-t-SNE and Neg-t-SNE~\cite{damrich2022t}. This selection allows us to assess CosMAP in relation to methods that are commonly used in practice and that address comparable challenges in neighbourhood preservation, cluster organization, and visual interpretability. To promote fairness and reproducibility, all methods are first compared using
their default parameter settings, without method-specific or dataset-specific
hyperparameter tuning. Further details on the default parameter values of
CosMAP, as well as its sensitivity analysis, are provided in the supplementary
material in Appendix~\ref{appendix_b}.
 An exception is made for particularly challenging datasets,
such as the kinship relatedness data, where CosMAP and the competing methods are
all tuned under the same evaluation protocol. This comparison allows us to assess
CosMAP within the same methodological family, whose main objective is to produce
interpretable two-dimensional representations while balancing local neighborhood
preservation and global organization.

\subsection{Handwritten-digit benchmarks: MNIST and USPS}
\label{sec:digit_benchmarks}

We first evaluate CosMAP on two standard handwritten-digit benchmarks, MNIST~\cite{Yan_lecun_mnist} and USPS~\cite{hull_1994}. These datasets provide a controlled setting for comparing dimensionality-reduction methods because the underlying classes are known and correspond to semantically meaningful digit categories from \(0\) to \(9\). Although the methods are applied in a fully unsupervised setting, the class labels can be used a posteriori to assess whether the resulting two-dimensional embeddings recover digit-wise structure. This makes MNIST and USPS useful benchmarks for evaluating the ability of each method to preserve neighborhood relationships and produce interpretable visual clusters.

\begin{figure*}[h!]
	\centering
	\includegraphics[width=0.6\textwidth]{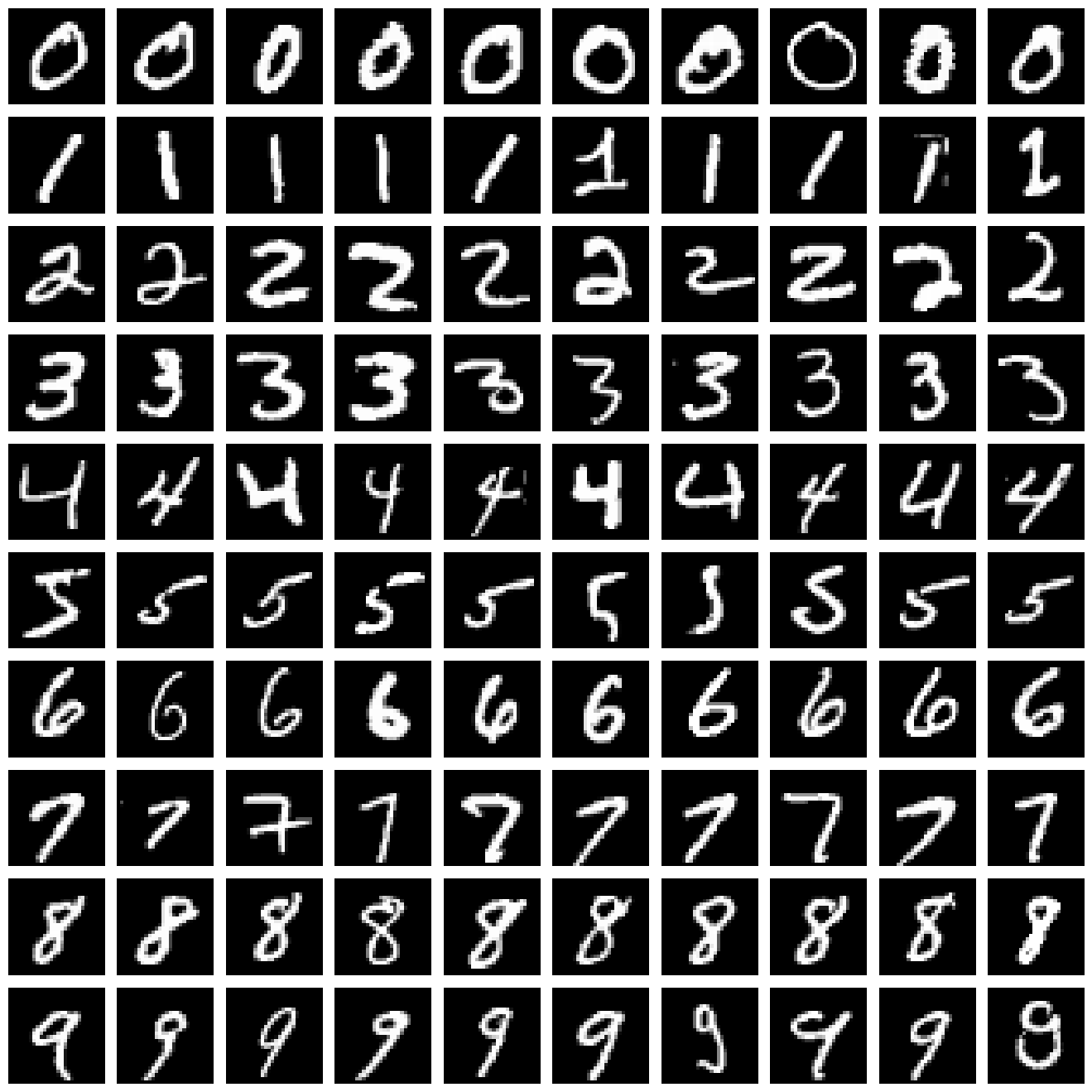}
	\caption{Ten representative MNIST images for each digit from 0 to 9.}
	\label{fig:mnist_examples_full}
\end{figure*}

MNIST contains \(70{,}000\) grayscale images of handwritten digits, each represented as a flattened \(28 \times 28\) image, yielding an input matrix \(X \in \mathbb{R}^{70000 \times 784}\). In our experiments, the dataset was loaded from OpenML using \texttt{fetch\_openml} with \texttt{as\_frame=False}. The corresponding label vector \(y\) contains the digit classes and is used only for visualization and post hoc evaluation, never during the optimization of the embeddings. No additional normalization, standardization, or train--test split was applied at this stage; the input features correspond to the original pixel-intensity values. Representative MNIST examples are shown in Figure~\ref{fig:mnist_examples_full}.

USPS is a complementary handwritten-digit dataset constructed from digit images scanned from envelopes by the U.S. Postal Service~\cite{hull_1994}. Each sample represents a digit from \(0\) to \(9\) and is encoded as a \(16 \times 16\) grayscale image, corresponding to a 256-dimensional input vector. Compared with MNIST, USPS is smaller and differs in acquisition and preprocessing, since the original digit images were normalized, deslanted, and resized. Therefore, using both MNIST and USPS allows us to evaluate whether the observed behavior of CosMAP is consistent across two related but distinct handwritten-digit datasets.

\begin{figure*}[h!]
	\centering
	\includegraphics[width=\textwidth]{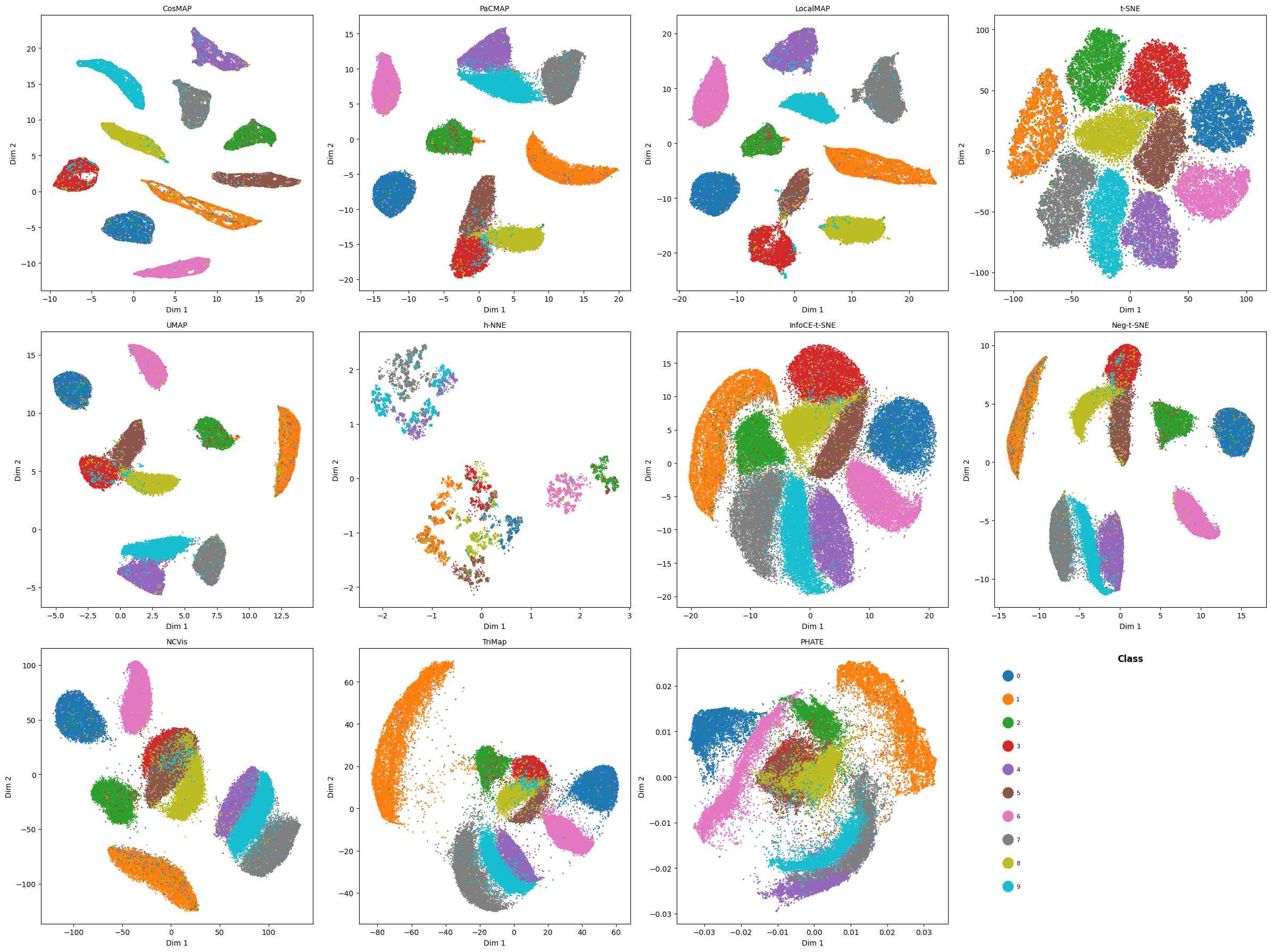}
	\caption{Comparison of dimensionality-reduction methods on the MNIST dataset.}
	\label{fig:cosmap-vs-sota-mnist}
\end{figure*}

\begin{figure*}[h!]
	\centering
	\includegraphics[width=\textwidth]{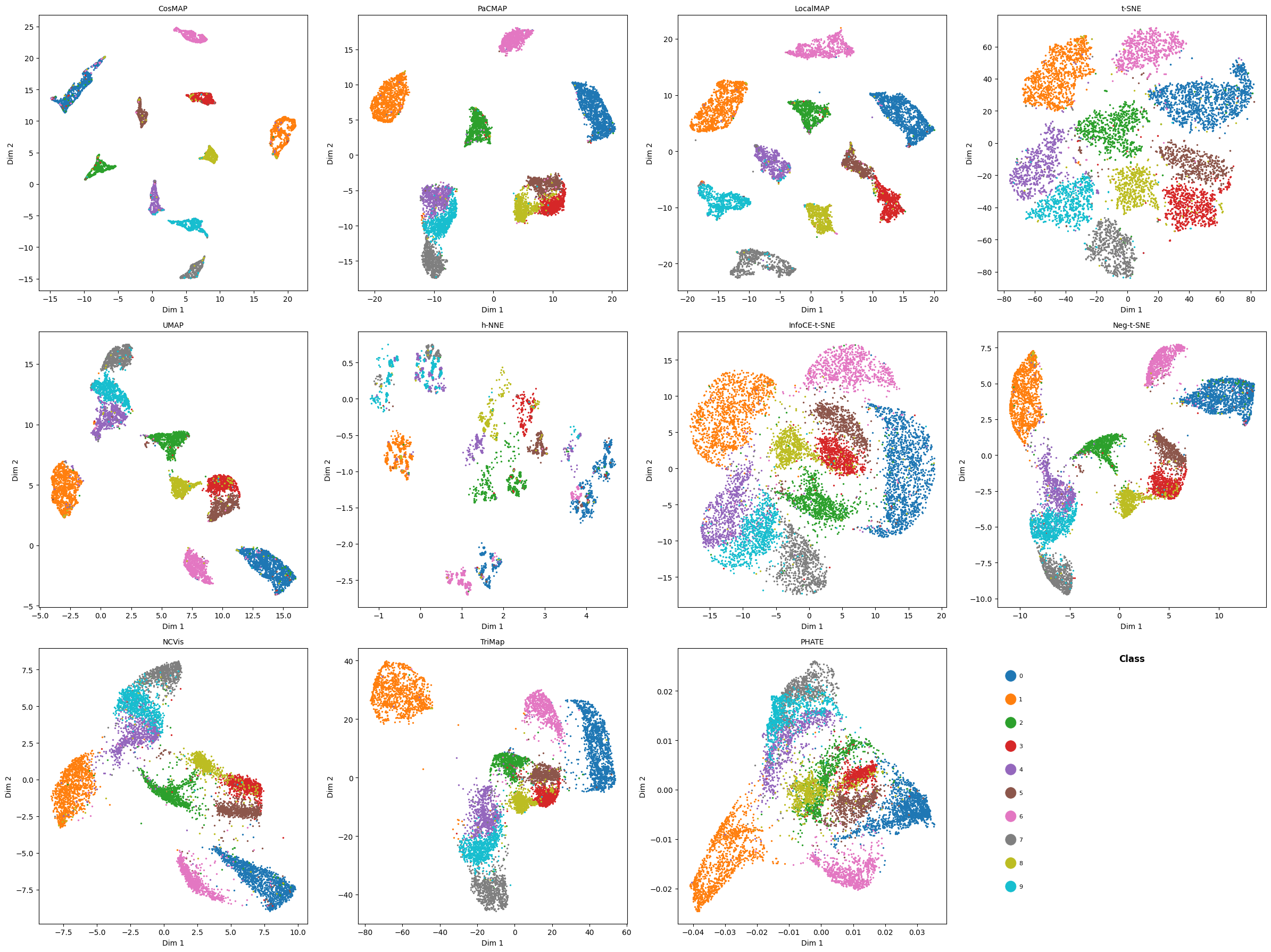}
	\caption{Comparison of dimensionality-reduction methods on the USPS dataset.}
	\label{fig:cosmap-vs-sota_usps}
\end{figure*}

Figures~\ref{fig:cosmap-vs-sota-mnist} and~\ref{fig:cosmap-vs-sota_usps} show the two-dimensional embeddings obtained on MNIST and USPS, respectively. Across both benchmarks, CosMAP produces the most visually interpretable cluster organization among the evaluated methods. The digit classes form compact and well-separated groups, suggesting that CosMAP preserves class-discriminative neighborhood structure while reducing overlap between visually similar digits. LocalMAP provides the closest visual performance to CosMAP and can be considered the second strongest method in these experiments, as it separates most digit classes effectively and yields a coherent global organization.

\begin{figure*}[h!]
    \centering

    \begin{subfigure}{0.48\textwidth}
        \centering
        \includegraphics[width=\textwidth]{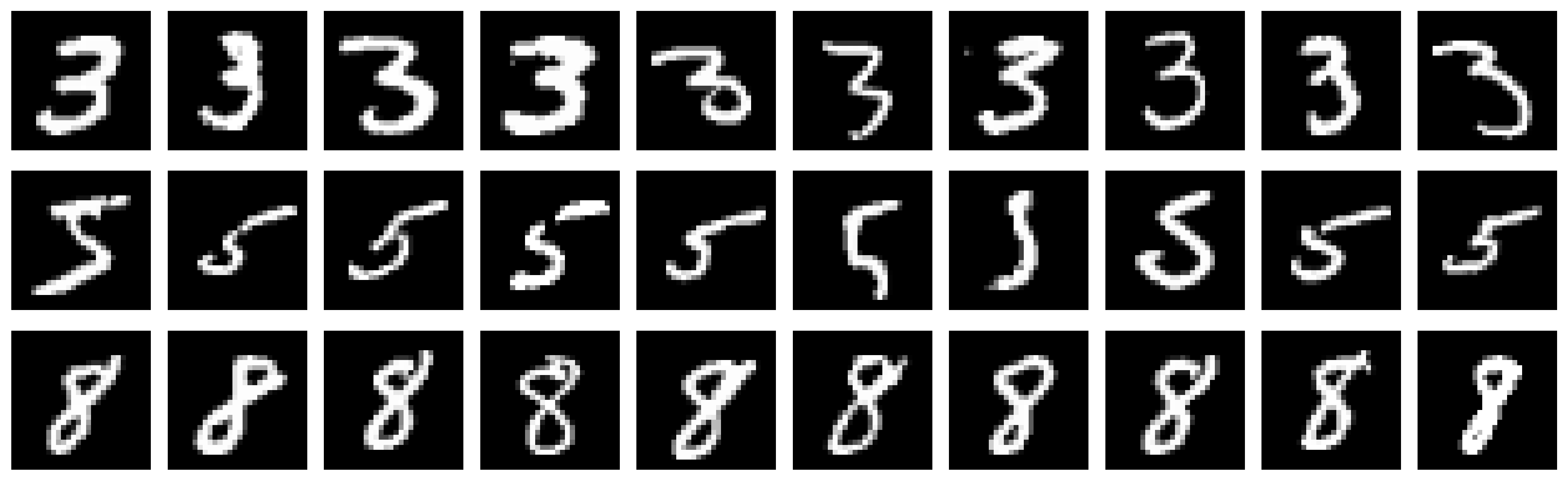}
        \caption{Examples of digits 3, 5, and 8.}
        \label{fig:mnist_examples_3_5_8}
    \end{subfigure}
    \hfill
    \begin{subfigure}{0.48\textwidth}
        \centering
        \includegraphics[width=\textwidth]{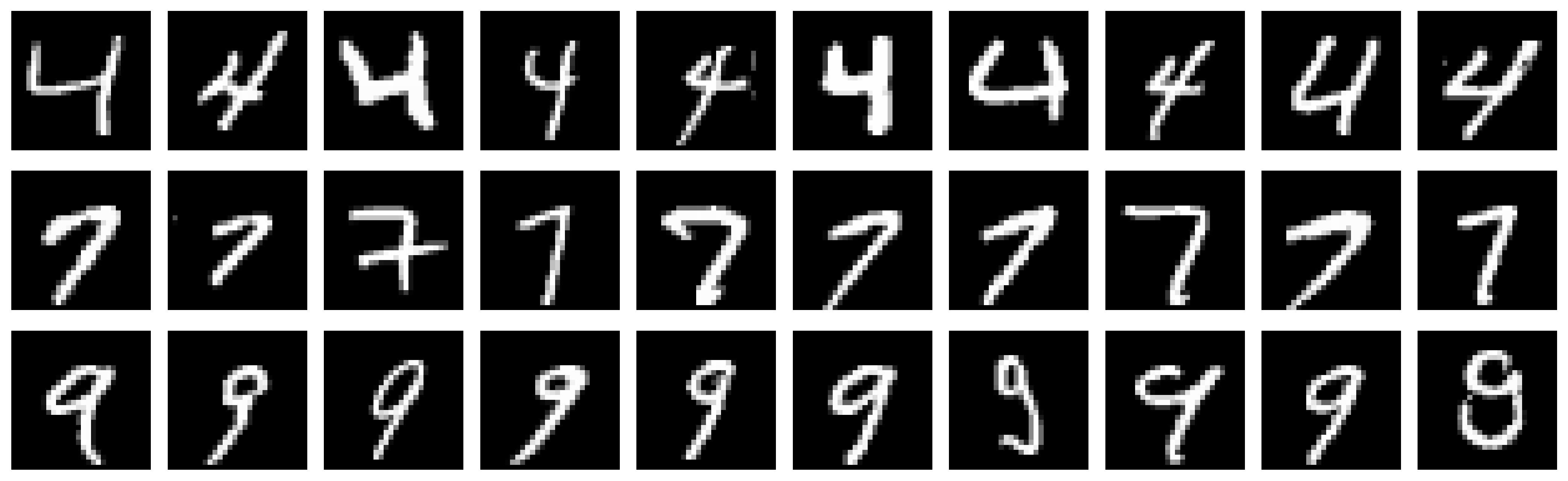}
        \caption{Examples of digits 4, 7, and 9.}
        \label{fig:mnist_examples_4_7_9}
    \end{subfigure}

    \caption{Examples of visually similar handwritten digits in MNIST, with 10 samples per class.}
    \label{fig:mnist_examples_side_by_side}
\end{figure*}

On MNIST, CosMAP yields a clear separation of the ten digit classes, with compact clusters and limited overlap. This behavior is particularly important for visually similar digits. For example, digits \(3\), \(5\), and \(8\) may share curved strokes and partial closures, while digits \(4\), \(7\), and \(9\) may exhibit similar angular structures or elongated strokes, as illustrated in Figure~\ref{fig:mnist_examples_side_by_side}. Several competing methods, including UMAP, PaCMAP, and Neg-t-SNE, recover meaningful digit-wise structure, but some of these ambiguous groups remain close or partially mixed. In contrast, CosMAP provides a clearer spatial separation of these difficult classes. LocalMAP also performs strongly, although some confusion remains between digits \(3\) and \(5\).

The USPS results confirm the behavior observed on MNIST. Despite the lower image resolution and different preprocessing of USPS, CosMAP again produces the clearest and most interpretable embedding. Most digit classes form compact groups, and the global layout remains readable without relying exclusively on color labels. LocalMAP remains the closest competing method, but it shows a more visible overlap between digits \(3\) and \(5\). UMAP and PaCMAP recover meaningful structures but present more overlap between certain classes, while t-SNE produces locally coherent clusters with a less interpretable global arrangement. Other methods, including NCVis, TriMAP, PHATE, h-NNE, InfoNCE-t-SNE, and Neg-t-SNE, show weaker separation or more fragmented embeddings in this setting.

Overall, the joint analysis of MNIST and USPS shows that CosMAP is highly competitive on handwritten-digit data. In both benchmarks, CosMAP provides the most interpretable two-dimensional representation, followed by LocalMAP as the strongest competing baseline. These results support the ability of CosMAP to preserve meaningful neighborhood structure and to produce compact, separated, and visually interpretable clusters across related image datasets with different resolutions and preprocessing characteristics.

\paragraph{Single-cell RNA-seq preprocessing.} \label{sec:scrna_preprocessing}

All single-cell RNA-seq datasets used in this work were processed with the
same preprocessing workflow in order to ensure comparability across methods and
datasets. These public real dataset were loaded as \texttt{AnnData} objects,  directly through the \texttt{scvi-tools} dataset API  
\cite{wolf2018scanpy,gayoso_scvi_tools_2021}. For each dataset, we first applied
a light gene-filtering step by removing genes with fewer than three total counts
across all cells \cite{stuart2019_seuratv3}. We then performed library-size normalization to a total of \(10^4\) counts per
cell, followed by a \(\log(1+x)\) transformation, where $\log$
 denotes the natural logarithm. To reduce the
feature dimension before applying the dimensionality-reduction methods, we
selected the \(1{,}200\) most highly variable genes using the
\texttt{"seurat\_v3"} option implemented in \texttt{Scanpy}
\cite{stuart2019_seuratv3,wolf2018scanpy}. This number of genes lies within the
commonly used range for single-cell RNA-seq preprocessing and provides a moderate
feature space for comparing dimensionality-reduction methods
\cite{luecken2019current,savae_2023}. Unless otherwise stated, CosMAP and all
baseline methods were applied to the same \(1{,}200\)-gene log-normalized
expression matrix.

This shared preprocessing protocol is used for all scRNA-seq benchmarks reported
in this work, including the Retina, Cortex, PBMC, and Heart Cell Atlas datasets.
The following subsections therefore describe only the origin, biological context,
and annotation structure of each dataset.

\subsection{Retina dataset}
\label{sec:retina_dataset}

The Retina dataset contains single-cell RNA-seq profiles of mouse retinal cell populations, including bipolar cells, which act as intermediate neurons between photoreceptors and ganglion cells. Because of its sparse transcriptomic nature and well-defined biological cell-type annotations, this dataset is useful for assessing whether dimensionality-reduction methods can preserve and reveal biologically meaningful cellular structures \cite{savae_2023}. The dataset has a size  $19829 \times 1200$ where the cell-type annotations are used only for visualization and evaluation;
they are not used during the optimization of CosMAP or any baseline method.

\begin{figure*}[h!]
	\centering
	\includegraphics[width=\textwidth]{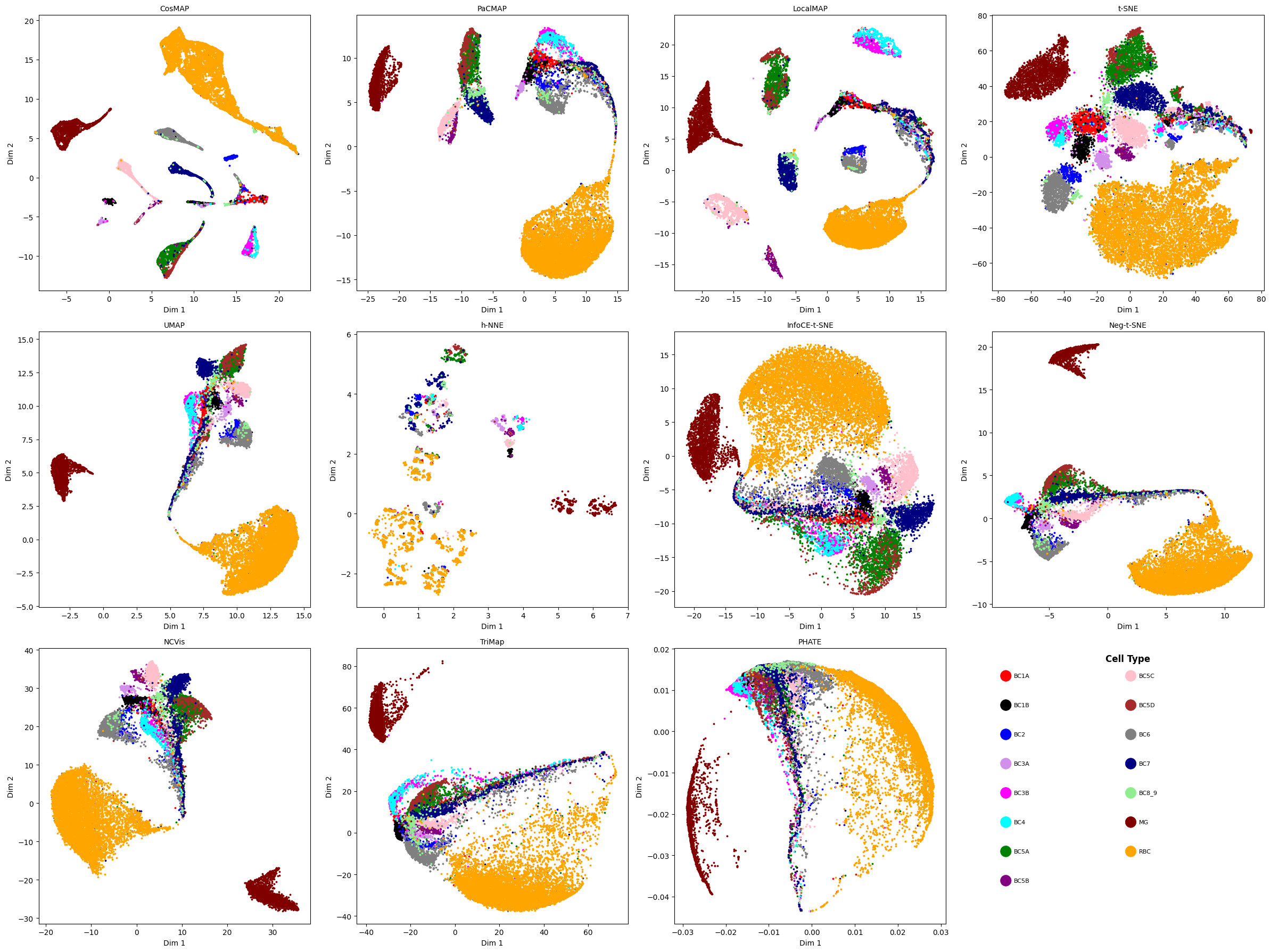}
	\caption{Comparison of DR methods on the Retina dataset.}
	\label{fig:cosmap-vs-sota-retina}
\end{figure*}

In this section, we qualitatively assess the embeddings produced by CosMAP, focusing on how well the method recovers known cell-type structure. Again, we compare CosMAP with the same standard DR methods  using two-dimensional embeddings, in order to  visually inspect cluster separation, continuity of developmental trajectories and the stability of local neighbourhoods across methods. Illustrative examples are shown in Figure ~\ref{fig:cosmap-vs-sota-retina}, where each panel highlights how CosMAP affects the geometry of the resulting embeddings. Here, several methods recover part of the global organization of the retina data, but with different limitations. t-SNE is one of the few baseline methods that clearly separates the  rod bipolar cell (RBC) cluster in orange from the main group. However, the remaining cell types are compressed into a crowded region, which makes the separation between smaller subtypes difficult to interpret. LocalMAP also produces an interesting class organization and separates several retinal populations more clearly than many other methods. Nevertheless, some clusters remain attached to, or partially connected with, the RBC region, suggesting that the distinction between the dominant RBC population and neighbouring bipolar-cell subtypes is not fully resolved.

Other methods, such as UMAP, PaCMAP, TriMAP, PHATE, h-NNE, NCVis, InfoCE-t-SNE, and Neg-t-SNE, show different compromises between global organization and local class separation. In several cases, the RBC population is either only partially detached from the main structure  such BC1A, BC2A, BC3B, etc., or the smaller classes are mixed together in a dense central region. This makes their embedding less informative for identifying fine-grained retinal subtypes. 
In this Retina dataset, the partial overlap observed between BC3B and BC4 should not be interpreted as a failure of CosMAP, but rather as a biologically plausible limitation of two-dimensional visualization for closely related retinal bipolar-cell subtypes. BC3A, BC3B, and BC4 belong to the OFF cone bipolar-cell family, together with BC1A, BC1B, and BC2. These OFF bipolar subtypes are known to be transcriptionally and anatomically close, which makes their complete separation difficult in a low-dimensional embedding. This interpretation is supported by Ding et al. \cite{ding2018scvis}, who projected an independent whole-retina dataset onto a bipolar-cell reference and reported that one cluster corresponded jointly to BC3B and BC4, while other clusters similarly grouped nearby subtypes such as BC2/BC3A and BC1A/BC1B \cite{ding2018scvis}. Therefore, the persistence of BC3B--BC4 or BC1A--BC1B mixing in both CosMAP and LocalMAP is very consistent with previous single-cell analyses rather than being an isolated artifact of the proposed method.

More importantly, CosMAP still reveals a meaningful improvement: while most methods fail to separate BC3A and BC3B from the neighboring OFF bipolar-cell group, CosMAP recovers a finer subdivision in which BC3A/BC3B-related structure becomes partially visible. The remaining BC3B--BC4 or  BC1A--BC1B overlap is expected because these two subtypes share a strong OFF bipolar transcriptional background. For example, Grik1 has been reported as a marker of BC2, BC3A, BC3B, and BC4, and Probe-Seq experiments showed that markers of these subtypes are enriched together in the Grik1-positive population \cite{amamoto2019probeseq}. In addition, morphological studies describe OFF bipolar cells as a sequence of closely related types stratifying in the outer inner plexiform layer, with type 3b and type 4 occupying adjacent positions in this organization \cite{tsukamoto2017classification}. Thus, the BC3B--BC4 overlap reflects the intrinsic similarity of these populations.

Finally, this behavior should be interpreted in light of the sparse and noisy nature of scRNA-seq data. Dropout and low RNA capture rates can obscure the marker genes that distinguish closely related cell types, making fine subtype boundaries difficult to preserve in two dimensions \cite{eraslan2019dca}. Consequently, the fact that CosMAP separates BC3A from part of the surrounding OFF bipolar-cell structure, while BC3B and BC4 remain partially mixed, supports the conclusion that CosMAP captures biologically meaningful fine-grained structure without artificially forcing a complete separation where the transcriptomic boundary is weak. By comparison, CosMAP provides a clearer separation of the  RBC cluster from the main population while maintaining a more readable organization of the remaining retinal cell types. This behaviour is also  consistent with the objective of similarity-assisted VAE \cite{savae_2023}, where the quality of the embedding is assessed not only by the separation of a dominant cell type but also by the preservation of interpretable groupings among smaller and closely related cell populations.

To quantitatively compare CosMAP with the baseline dimensionality-reduction methods, we assess the agreement between the known cell-type annotations and the cluster structure obtained from each two-dimensional embedding. We use Normalized Mutual Information (NMI), a standard information-theoretic measure for comparing two partitions of the same set of observations \cite{strehl2002cluster}. NMI has also been used to assess the clustering quality of representations produced by dimensionality-reduction methods, particularly in the context of nonlinear data \cite{strehl2002cluster, Roh_hae_Clustering2025}. Let $Y$ denote the ground-truth cell-type labels and let $C$ denote the cluster assignments obtained by applying $k$-means to the two-dimensional embedding, with the number of clusters fixed to the number of annotated classes. The mutual information between $Y$ and $C$ is defined as
\begin{equation}
I(Y;C)
= \sum_{y} \sum_{c}
p(y,c) \log \frac{p(y,c)}{p(y)p(c)},
\end{equation}
where p$(y,c)$ is the empirical joint distribution of labels and clusters, while $p(y)$ and $p(c)$ are the corresponding marginal distributions. We then report the normalized form
\begin{equation}
\mathrm{NMI}(Y,C)
= \frac{2I(Y;C)}{H(Y) + H(C)},
\end{equation}

where
\begin{equation}
H(Y) = -\sum_{y} p(y)\log p(y), \qquad
H(C) = -\sum_{c} p(c)\log p(c).
\end{equation}
This normalization gives values in the interval \([0,1]\), where higher values indicate stronger agreement between the clusters recovered from the embedding and the annotated cell types. For each two-dimensional embedding, we applied $k$-means clustering with the number of clusters $k$ fixed to the number of unique cell-type labels available in the corresponding dataset\footnote{The labels are used only after the embeddings have been obtained, for quantitative evaluation purposes.}. We then computed the NMI between the cluster assignments produced by $k$-means and the reference cell-type annotations. Both the $k$-means algorithm and the NMI score were implemented using the \texttt{scikit-learn} library \cite{pedregosaScikitlear_2018}. Because $k$-means is sensitive to the random initialization of its centroids and may converge to different local optima, we performed five independent runs with different random seeds for each embedding. This repeated evaluation reduces the influence of any single favorable or unfavorable initialization and provides a more robust assessment of the cluster structure preserved by each dimensionality-reduction method.

\begin{figure*}[h!]
	\centering
	\includegraphics[width=\textwidth]{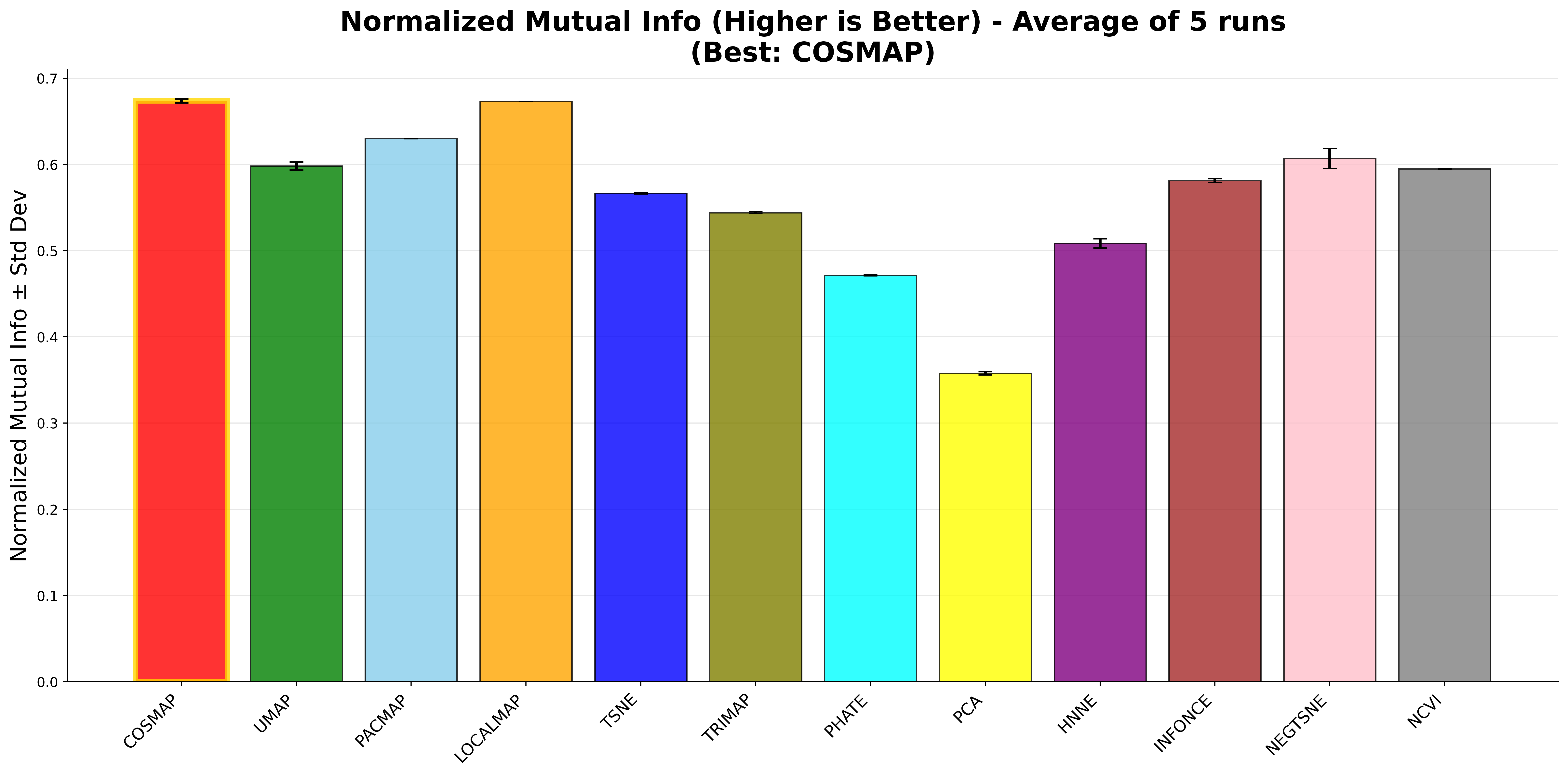}
	\caption{Quantitative evaluation DR methods on the Retina dataset.}
	\label{fig:metric-retina-nmi}
\end{figure*}
Figure~\ref{fig:metric-retina-nmi} reports the average NMI scores obtained over five independent runs on the Retina dataset. CosMAP achieves the highest average NMI among the evaluated dimensionality-reduction methods, indicating that the clusters obtained from its two-dimensional embedding show the strongest agreement with the annotated cell-type labels. This quantitative result supports the visual observations reported above: CosMAP produces a Retina embedding in which the main cell populations are clearly organized and well aligned with the known biological annotations. 


\subsection{Cortex dataset}
\label{sec:cortex_dataset}

The Cortex dataset contains mouse cortical cells spanning several neuronal and
glial populations \cite{zeisel2015cortex}. It provides a relatively small and
well-annotated benchmark for assessing whether dimensionality-reduction methods
can preserve known cellular organization in neural tissue. After applying the same preprocessing protocol described in Section~\ref{sec:scrna_preprocessing}, the resulting data has size $3005 \times 1200$. This processed data was then used as input for CosMAP and all baseline methods.

\begin{figure*}[h!]
	\centering
	\includegraphics[width=\textwidth]{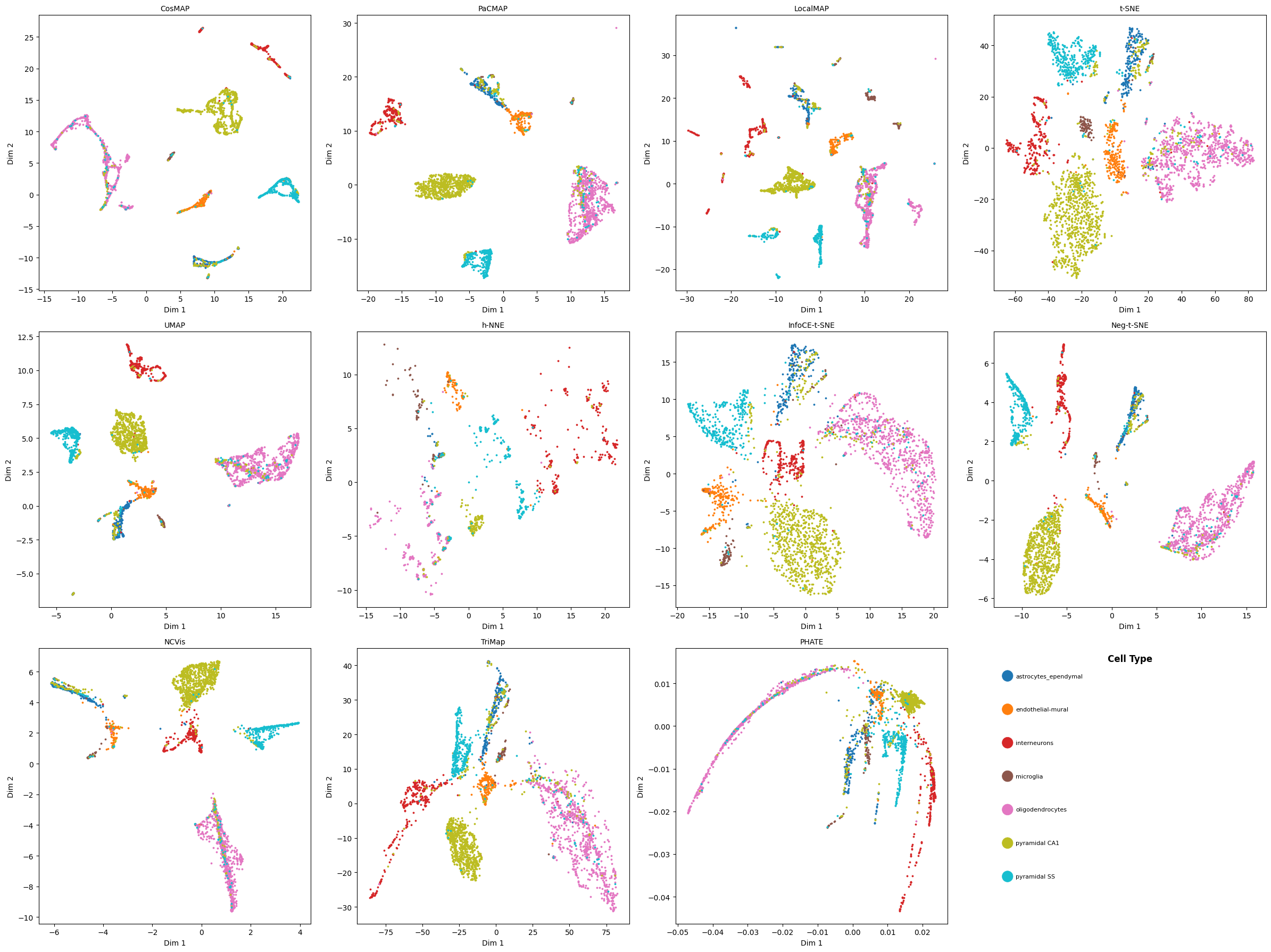}
	\caption{Comparison of DR methods on the mouse cortex dataset.}
	\label{fig:cosmap-vs-sota-cortex}
\end{figure*}
This dataset contains several biologically distinct but transcriptionally related cell types, including astrocyte/ependymal cells, endothelial-mural cells, interneurons, oligodendrocytes, and pyramidal neurons. Therefore, a good visualization should separate the main annotated populations while avoiding excessive fragmentation of the same cell type.

In Figure~\ref{fig:cosmap-vs-sota-cortex}, CosMAP provides a competitive visualization with well-separated cell-type clusters. In particular, the astrocyte/ependymal population, shown in blue, and the endothelial-mural population, shown in orange, are more clearly distinguished than in several other methods, where these two groups tend to remain close, partially mixed, or attached to neighbouring populations. This separation is important because small or less abundant cell populations are often difficult to isolate in two-dimensional embeddings. Among the baseline methods, LocalMAP produces an interesting visualization and separates several classes more strongly than UMAP, PaCMAP, PHATE, PCA, h-NNE, NCVis, InfoCE-t-SNE. However, t-SNE and its variant Neg-t-SNE achieve better visual separation. Notably, Neg-t-SNE produces visualizations similar to our method, possibly because both approaches rely on a comparable negative sampling strategy. One limitation visible in this dataset is that the local graph adjustment such as LocalMAP can also over-separate some populations. For example, the interneuron population, shown in red, as well as some  pyramidal populations, appear split into several disconnected fragments. This suggests that LocalMAP may reveal fine local structure, but can also fragment a biologically coherent class into multiple islands. CosMAP is not completely free from this issue. In particular, the endothelial-mural population in orange is separated into two subclusters, suggesting that the method may sometimes over-refine the embedding. This effect may be related to the small size of the cortex dataset, where the two-phase refinement can become too strong and amplify local differences within the same annotated class. Thus, although refinement improves separation in several datasets, it is not always optimal for small datasets with few cells and subtle class boundaries. Overall, CosMAP remains competitive on the cortex dataset. It provides clearer separation of difficult populations such as astrocyte/ependymal and endothelial-mural cells, while maintaining a generally readable organization of the remaining cell types. 
\begin{figure*}[h!]
	\centering
	\includegraphics[width=\textwidth]{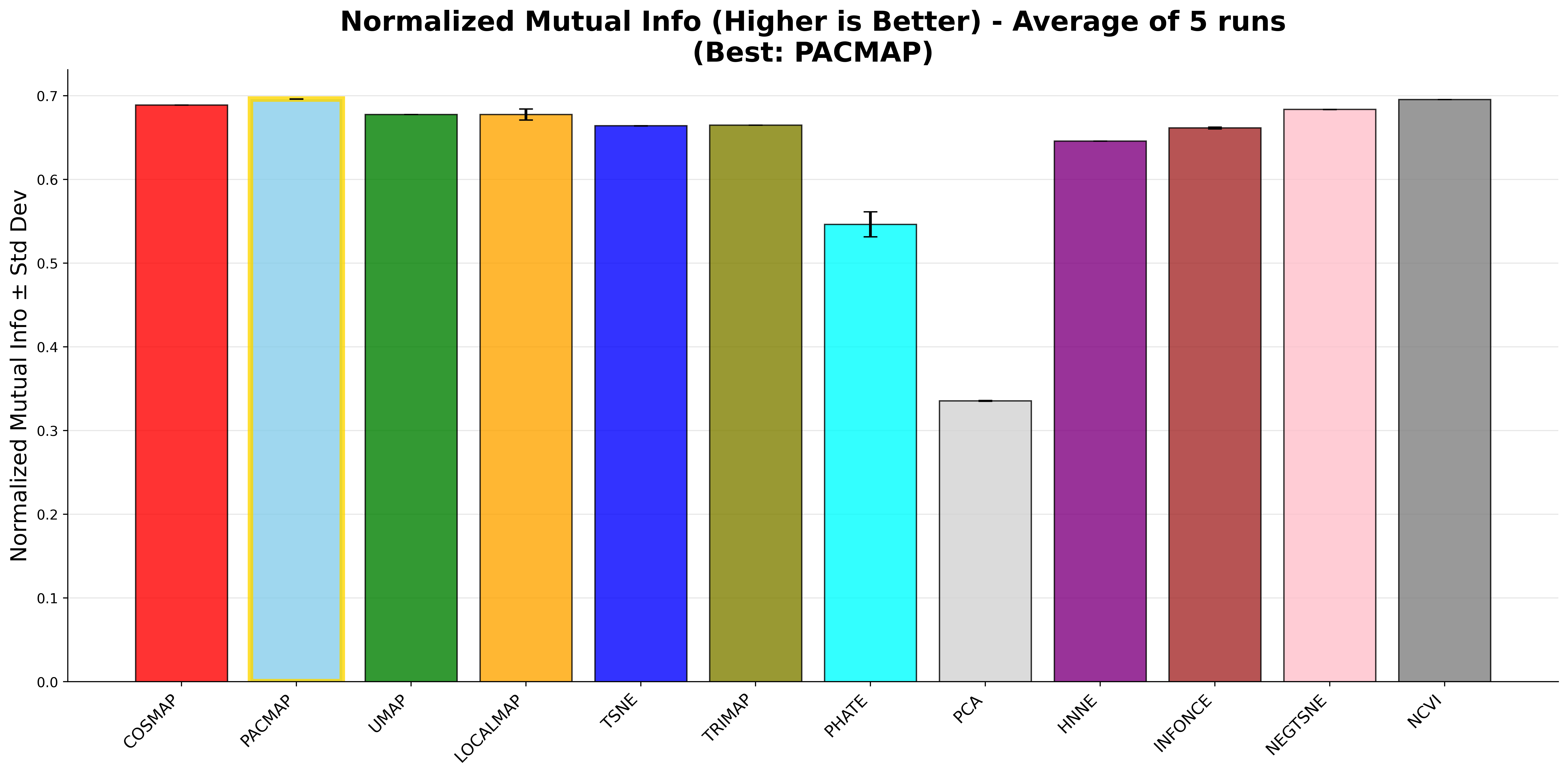}
	\caption{Quantitative evaluation DR methods on the Mouse cortex dataset.}
	\label{fig:metric-cortex-nmi}
\end{figure*}

For the quantitative evaluation on  the Cortex dataset in figure ~\ref{fig:metric-cortex-nmi}.  PaCMAP obtains the highest NMI score, followed by NCVis and CosMAP. Although CosMAP does not achieve the best score on this dataset, it remains among the top-performing methods, which indicates that its embedding still recovers a cluster structure that is well aligned with the annotated cell-type labels.

\subsection{Genealogical Dataset}
\noindent
To assess CosMAP's ability to uncover population structure from a genealogical dataset, we apply it to a pairwise kinship matrix derived from the  BALSAC--CARTaGENE dataset \citep{vezina_2018}. Let $\boldsymbol{\Phi}=\phi_{ij}\in\mathbb{R}^{n\times n}$ denote the resulting symmetric matrix, where each entry $\phi_{ij}$ quantifies the expected genealogical relatedness between individuals $i$ and $j$. More precisely, the kinship coefficient is defined as the probability that two alleles randomly sampled at the same locus, one from each individual, are identical by descent (IBD), meaning that they originate from a common ancestral allele \citep{kirkpatrick2019efficient,morin2026finescale}.

Consequently, in contrast to conventional feature-based datasets, the input is
itself a pairwise relational matrix, in which larger entries denote stronger
genealogical proximity. The pairwise kinship coefficients were computed with
GeneaKit \cite{morin2026finescale}, a Python package for pedigree analysis tailored to genealogical data
and built upon the functionality of GENLIB \cite{gauvinGENLIB_2015, morin2026finescale}. Following the processing strategy of Morin et al.\ \cite{morin2026finescale}, we obtained a final kinship matrix of size \(7878 \times 7878\). The regional labels were used exclusively to visualize the resulting embeddings and were not incorporated at any stage of the CosMAP optimization.

\begin{figure*}[h!]
	\centering
	\includegraphics[width=1.15\textwidth]{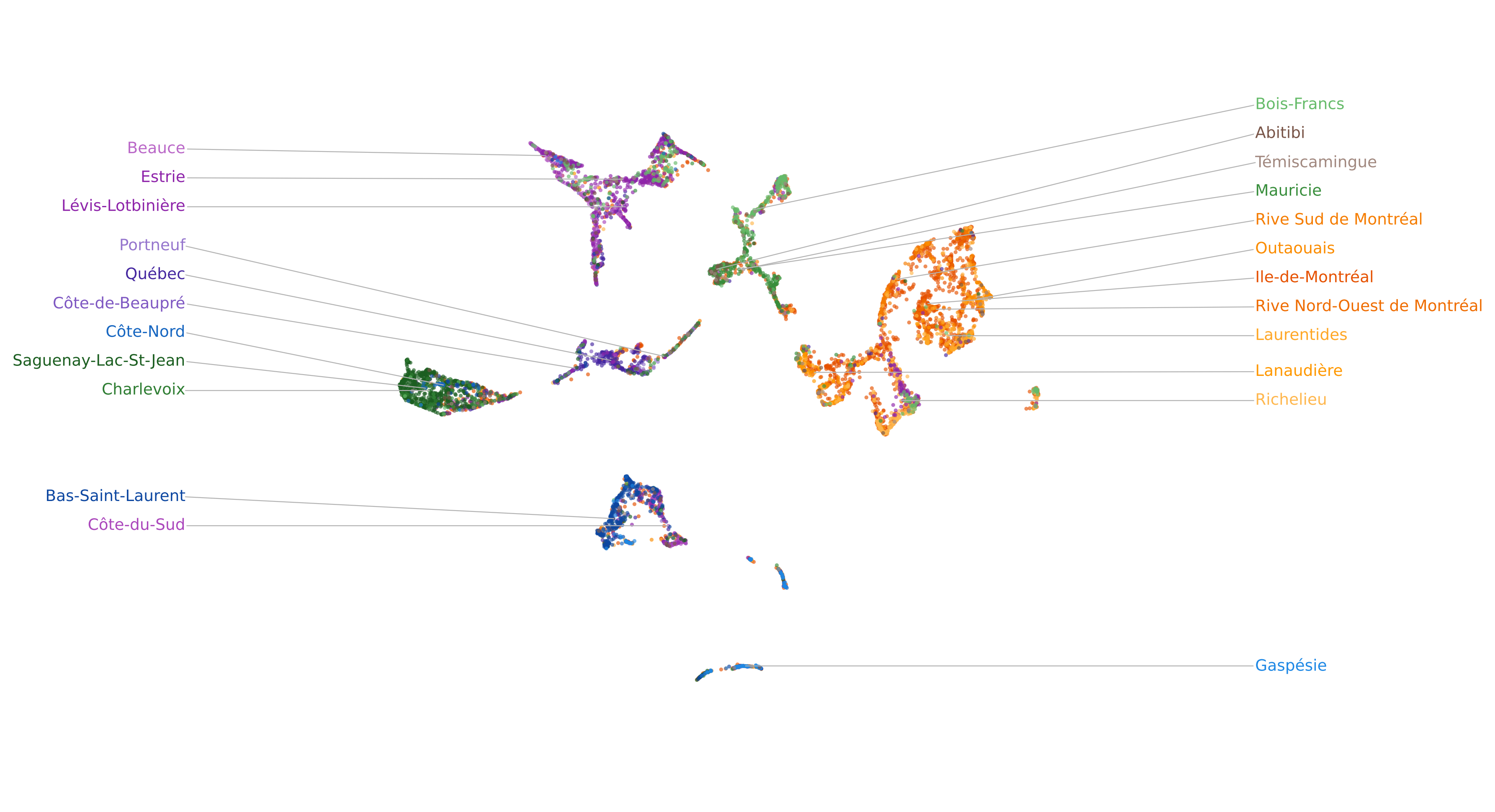}
	\caption{CosMAP embedding of the Cartagène kinship dataset. Each point represents one individual, and colors indicate the Quebec region in which the individual's parents were married.}
	\label{fig:cosmap-vs-sota-annotated-cosmap}
\end{figure*}

\noindent
A note on the experimental setup is needed before comparing the methods. In our
case the input is not a set of feature vectors but a square matrix of pairwise
similarities, and most dimensionality-reduction methods are not designed for this
kind of input. As a result, they tend to perform poorly when run with their
default settings. We therefore carried out a per-method hyperparameter search, selecting
for each technique the configuration that produced the most informative
two-dimensional embedding. A perfectly uniform comparison is nonetheless
infeasible, since the methods do not expose an identical interface. As an example: UMAP and
CosMAP accept a \emph{precomputed} metric, whereas several baselines do not. To
make the comparison as fair as the methods permit, we fixed the Euclidean
distance as the working metric for this dataset, the single choice supported by
all methods considered, and held the number of neighbours and the remaining
shared hyperparameters constant across techniques. This ensures that observed
differences in the embeddings reflect the methods themselves rather than
divergent metric or neighbourhood settings.

\begin{figure*}[h!] \centering \includegraphics[width=\textwidth] {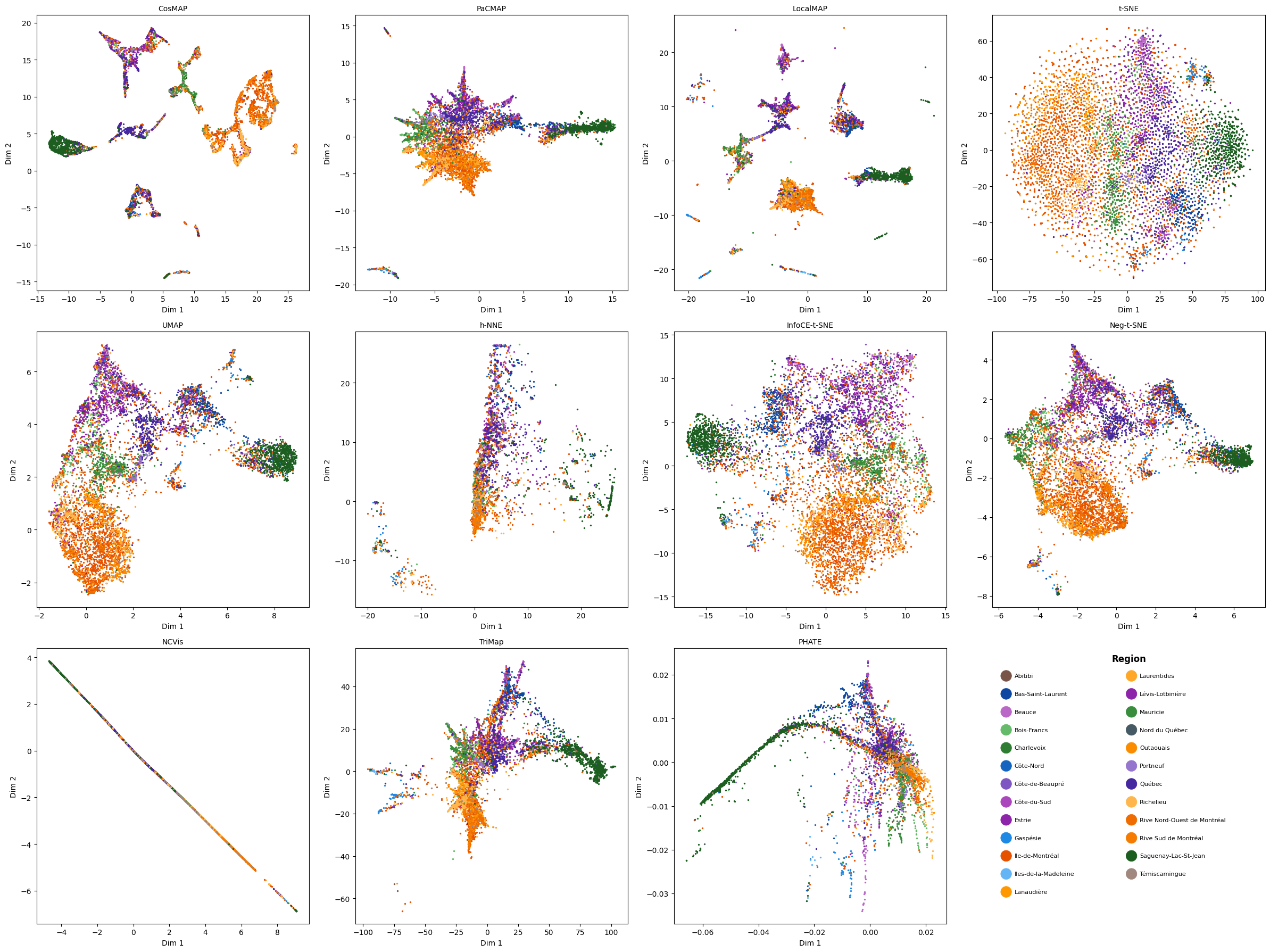} \caption{Comparison of dimensionality-reduction methods on the CARTaGENE kinship matrix. Each point represents one individual, and colors indicate the Quebec region in which the individual's parents were married.} \label{fig:cosmap-vs-sota-kinship} \end{figure*}

\noindent Figure~\ref{fig:cosmap-vs-sota-kinship} compares the two-dimensional embeddings obtained by applying CosMAP and the baseline dimensionality-reduction methods to the processed kinship matrix. Each point represents one individual. The regional label associated with an individual corresponds to the location, expressed as a Quebec administrative region, where that individual's parents were married. All of these baseline methods are unsupervised, so  these labels were not used during the construction or optimization of any embedding and are displayed only for the purpose of interpretation.

\noindent To facilitate visual comparison, regions were grouped into related colour families. Greater Montr\'eal and its surrounding regions are represented using orange tones, whereas Qu\'ebec City and its surrounding regions are shown in purple. Saguenay--Lac-Saint-Jean and Charlevoix are represented in green; Bas-Saint-Laurent, C\^ote-Nord, Gasp\'esie, and \^Iles-de-la-Madeleine are included in the blue family; and the more remote northern or western regions are represented using brown and grey tones.
\cite{dionne2023quebec,morin2026finescale}.

\noindent The baseline methods recover the regional structure encoded in the kinship matrix to varying degrees. LocalMAP identifies some of the main large-scale groups, particularly the distinction between the orange Montr\'eal-associated regions and the green Saguenay--Lac-Saint-Jean/Charlevoix group. However, the Qu\'ebec-associated regions remain comparatively mixed, and several of the smaller regional structures are less clearly distinguishable. Other methods similarly produce partial regional organization, but with greater overlap between colour families or less coherent arrangements of individuals sharing the same broad regional background. \noindent This limitation is especially apparent for NCVis. In contrast to its behaviour on the previous datasets, NCVis does not recover a clearly interpretable two-dimensional structure for the kinship matrix. To determine whether this result was caused by a single unfavourable initialization, the experiment was repeated over five independent runs. A similar behaviour was observed across the runs, suggesting that the result is more likely related to the difficulty of representing or optimizing this particular genealogical relatedness matrix than to one isolated initialization. \noindent By comparison, the CosMAP result (in figure ~\ref{fig:cosmap-vs-sota-annotated-cosmap}) provides the clearest and most coherent visual organization of the broad regional structures. In the CosMAP embedding, the regions associated with Greater Montr\'eal form a large orange structure on the right-hand side, whereas Saguenay--Lac-Saint-Jean and Charlevoix emerge as a distinct green structure. Bas-Saint-Laurent, C\^ote-Nord, Gasp\'esie, and \^Iles-de-la-Madeleine appear as more peripheral blue groups. The regions associated with Qu\'ebec City, represented in purple, also remain visually distinguishable instead of being entirely absorbed into the Montr\'eal- or Saguenay-associated structures. Thus, among the methods considered in Figure~\ref{fig:cosmap-vs-sota-kinship}, CosMAP produces the most interpretable separation of the principal regional families while retaining a coherent global arrangement of the individuals. \noindent This organization is particularly relevant for the kinship dataset because the input matrix directly represents pairwise genealogical similarity. The CosMAP result suggests that the method preserves an important part of this relational structure in two dimensions while keeping the main regional patterns visually interpretable. Since the regional labels were introduced only after the embedding had been computed, the observed colour organization reflects structure already contained in the kinship matrix rather than information explicitly supplied to CosMAP. \noindent Nevertheless, the embedding should not be interpreted as a geographic map, and the observed groups should not be regarded as a perfect classification of Quebec administrative regions. Distances in the embedding represent similarity between kinship profiles rather than physical geographic distance. Individuals located close to one another therefore tend to exhibit similar patterns of genealogical relatedness. The correspondence observed between these patterns and the regions where their parents were married may reflect regional founder effects, historical migration, and the broader population structure of Quebec \cite{dionne2023quebec,morin2026finescale}. Overall, CosMAP reveals this global genealogical-geographic organization without using any explicit regional information during optimization.

\begin{figure*}[h!]
	\centering
	\includegraphics[width=\textwidth]{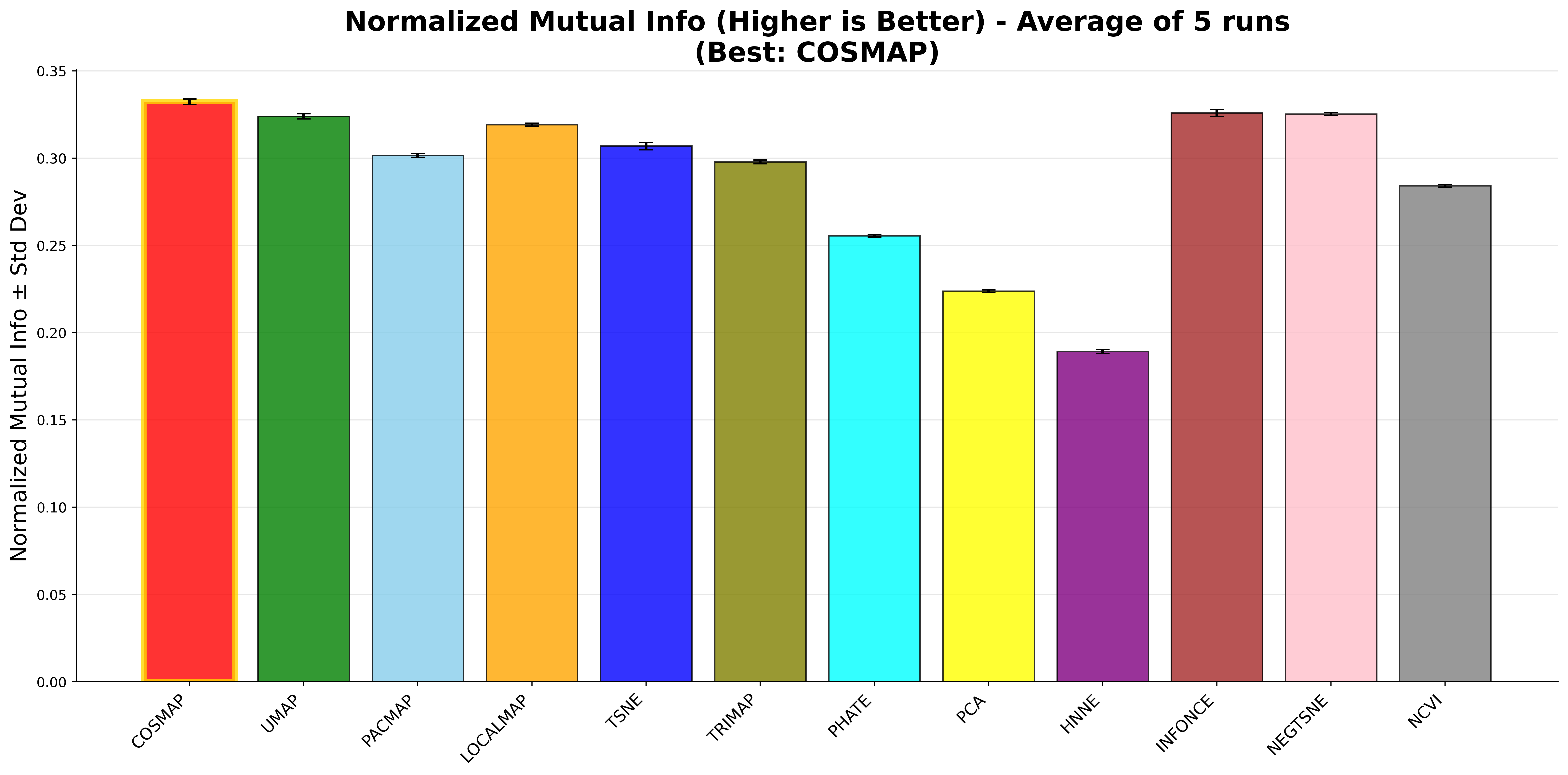}
	\caption{Quantitative evaluation DR methods on the Kinship dataset.}
	\label{fig:metric-Kinship-nmi}
\end{figure*}

For the quantitative evaluation on this kinship-relatedness matrix, CosMAP attains the best average NMI across five different runs, outperforming all other methods (Figure~\ref{fig:metric-Kinship-nmi}). This
indicates that the cluster structure recovered from the CosMAP embedding is more
consistent with the known regional labels than that obtained by the baselines.
The quantitative result corroborates the visual comparison and confirms that
CosMAP is particularly effective for this kinship-based representation.

\subsection{Software Implementation and Computational Framework}

CosMAP was implemented in Python, leveraging \texttt{PyTorch}~\cite{paszkePyTorchImperativeStyle2019} and \texttt{NumPy}~\cite{harris2020array} for efficient tensor operations and optimization of the embedding. The implementation is designed primarily for hardware-accelerated execution. When a CUDA-enabled GPU is available, both the construction of the $k$-nearest-neighbor graph and the optimization of the embedding are performed on the GPU. On Apple devices, acceleration may instead rely on the Metal Performance Shaders backend. Because these operations represent the main computational bottlenecks of the method, GPU execution is recommended, particularly for large datasets. To control memory usage during optimization, attractive and repulsive updates are evaluated in batches rather than by materializing all pairwise interactions simultaneously. This batch-based strategy makes it possible to benefit from GPU parallelism while limiting the amount of device memory required during training. For nearest-neighbor search, \texttt{FAISS}~\cite{douze2024faiss} can be used as the accelerated backend, especially when CUDA support is available. When no compatible accelerator is detected, CosMAP automatically falls back to a CPU-based execution pathway. In this case, approximate $k$-nearest-neighbor search is performed using \texttt{PyNNDescent}~\cite{EfficientKnearestNeighbor}, while the embedding updates are accelerated with \texttt{Numba}~\cite{lamNumba_2015}. Although this fallback allows the method to run on standard computing environments, it is generally slower than the GPU implementation and is therefore more suitable for moderate-sized datasets. After the neighborhood graph has been constructed, the high-dimensional affinity matrix \(P=(p_{ij})\) is stored and manipulated in sparse form using \texttt{SciPy}~\cite{virtanen2020scipy}. When spectral initialization is selected, the corresponding initialization procedure relies on routines provided by \texttt{umap-learn}~\cite{HowUMAPWorksHowUMAPWorksipynbMaster}. The experimental workflow was developed using standard scientific Python libraries. In particular, \texttt{pandas}~\cite{mckinneyPandas_2011} was used for data organization, \texttt{scikit-learn}~\cite{pedregosaScikitlear_2018} for preprocessing and quantitative evaluation, and \texttt{Matplotlib}~\cite{hunter2007matplotlib} for visualization. In addition, \texttt{GeneaKit} was used specifically to preprocess the kinship data and compute the pairwise kinship coefficients. The benchmark methods were executed using their respective Python implementations, including \texttt{phate}~\cite{phate_2019}, \texttt{umap-learn}~\cite{HowUMAPWorksHowUMAPWorksipynbMaster}, \texttt{hnne}~\cite{sarfraz2022hnne}, \texttt{contrastive-ne}~\cite{damrich2022t}, \texttt{trimap}~\cite{amid_trimap_2022}, \texttt{scikit-learn}~\cite{pedregosaScikitlear_2018}, and \texttt{pacmap}~\cite{wang2021pacmap}.

Additional implementation details concerning the main components of CosMAP, including optimization, initialization, and the two-phase refinement procedure, are provided in Appendix~\ref{appendix_a}. To facilitate reproducibility, reuse, and future extensions, the complete source code, software dependencies, and scripts used to generate the experimental results are publicly available in the CosMAP repository: \url{https://github.com/FenosoaRandrianjatovo/CosMAP-dr/blob/main/experiments/run_experiments.ipynb}.

\section{Conclusion and future direction}
\label{sec:conclusion_discussion}

In this work, we introduced CosMAP, a nonlinear unsupervised dimensionality-reduction method designed to produce interpretable low-dimensional representations of high-dimensional data. CosMAP combines a  high-dimensional affinity with a temperature-normalized contrastive formulation and a heavy-tailed low-dimensional similarity kernel to better preserve meaningful neighbourhood relationships while improving the visual organization of the embedding. Through experiments on handwritten-digit benchmarks, single-cell RNA sequencing datasets, and kinship data, we showed that CosMAP produces competitive and often highly interpretable two-dimensional embeddings compared with widely used visualization-oriented methods such as t-SNE~\cite{vandermaaten2008tsne}, UMAP~\cite{umap_2020}, PaCMAP~\cite{wang2021pacmap}, LocalMAP~\cite{wang2025localmap}, PHATE~\cite{moon2019phate}, TriMAP~\cite{amid_trimap_2022},  NCVis~\cite{artemenkov2022ncvis}, h-NNE~\cite{sarfraz2022hnne}, and InfoNCE-t-SNE/Neg-t-SNE~\cite{damrich2022t}. Overall, the empirical results suggest that CosMAP is particularly effective at producing compact and well-separated clusters while maintaining a readable global organization. On MNIST and USPS, CosMAP yielded the most interpretable  cluster structure among the evaluated methods, with LocalMAP providing the closest competing performance. On biological datasets, CosMAP also produced meaningful visualizations and showed strong ability to preserve  neighborhood structure, while reducing some of the over-fragmentation observed with certain competing methods. These results support the relevance of CosMAP as a visualization tool for exploratory analysis in settings where labels are unavailable or used only for post hoc interpretation. Although CosMAP shows competitive behaviour across the datasets considered in this study, it should not be interpreted as a universally optimal dimensionality-reduction method. As for other nonlinear dimensionality-reduction  methods, its performance may vary depending on the intrinsic structure of the data  and the chosen hyperparameters. One limitation observed in our experiments concerns the two-phase refinement strategy. In several datasets, learning an intermediate representation before the final two-dimensional projection improves the organization, separation, and readability of the embedding. Nevertheless, this refinement does not always provide a clear advantage over the direct two-dimensional optimization. Its effectiveness appears to depend on the intrinsic geometry of the dataset. For example, on the Cortex dataset, the full pipeline sometimes produces a stronger separation of small groups across different runs, whereas stopping after the first phase gives a more coherent representation. This limitation should not be interpreted as a specific weakness of CosMAP. Rather, it suggests that stronger local refinement is not universally optimal. A comparable behavior can also be observed with methods such as LocalMAP, which dynamically adjusts the neighborhood graph during optimization to recover clusters that may be merged by other methods \cite{wang2025localmap}. While this strategy is useful in many cases, it may also accentuate weak local variations and produce over-separated visual structures when the data are small, noisy, or when the original graph is already sufficiently informative. Therefore, both CosMAP refinement and locally adjusted graph strategies  in LocalMAP must be used with care.


In conclusion, CosMAP provides a competitive and interpretable framework for unsupervised visualization of high-dimensional data. While further methodological and empirical work is needed to improve its adaptivity and to better characterize its behaviour across data types, the results presented in this study suggest that CosMAP is a promising contribution to the family of nonlinear dimensionality-reduction methods for exploratory data analysis. Our results show that the optimal choice of metric is strongly dataset-dependent: cosine similarity is well suited to sparse, high-dimensional data such as scRNA-seq and image datasets, while Euclidean distance yields more meaningful representations for the Kinship CARTaGENE data. This underscores the critical role of the similarity measure in constructing the neighborhood graph for graph-based dimensionality reduction.  CosMAP currently supports Euclidean distance, cosine similarity, and precomputed similarity matrices. Future work will therefore focus on extending CosMAP to additional metrics and on developing practical criteria for selecting between one-phase and two-phase refinement. Finally, to ensure reproducibility and facilitate future use, CosMAP has been implemented as an open-source Python package compatible with the \texttt{scikit-learn} estimator API \cite{pedregosaScikitlear_2018}. This design allows the method to be easily integrated into standard machine-learning pipelines and single-cell data-analysis workflows. The source code is released under the BSD 2-Clause license, promoting transparency, reuse, and further methodological development. 

\noindent Our source code is available at \href{https://github.com/FenosoaRandrianjatovo/CosMAP-dr/tree/main/src/cosmapdr}{https://github.com/FenosoaRandrianjatovo/CosMAP-dr}.
\bibliographystyle{unsrt} 
\bibliography{references}
\newpage
\begin{center}
    \Huge\textbf{Appendix A}
\end{center}

\section{Implementation details of graph layout optimization in CosMAP}
\label{appendix_a}
\label{subsec:cosmap_optimization}

\noindent After constructing the weighted  graph in high dimension space, the embedding is optimized
directly in the low-dimensional space. Let
\[
E=\{(i,j):p_{ij}>0\}
\]
be the set of non-zero graph edges. In our implementation, these weights are internally
rescaled by their maximum value,
\[
\widetilde p_{ij}
=
\frac{p_{ij}}{\max_{(u,v)\in E} p_{uv}},
\]
so that \(\widetilde p_{ij}\in[0,1]\). 
This rescaling is only an implementation-level normalization used to keep the
magnitude of the attractive updates numerically stable. It preserves the
relative ordering of edge strengths and does not modify the topology of the
CosMAP graph. \cite{umap_2020, damrich2021umaploss}.

\noindent Given low-dimensional coordinates \(y_i\in\mathbb{R}^d\). The layout is optimized by minimizing a sampled binary cross-entropy objective. Positive graph edges contribute an attractive term, while randomly sampled non-neighbor pairs contribute a repulsive term:
\begin{equation}
\label{eq:sampled_bce}
    \widehat{\mathcal L}
=
-\sum_{(i,j)\in E_b}
\widetilde p_{ij}\log q_{ij}
-
\gamma
\sum_{(i,k)\in N_b}
\log(1-q_{ik}),
\end{equation}
where \(E_b\) is a mini-batch of positive graph edges, \(N_b\) is the set of negative pairs sampled uniformly from the vertices, and \(\gamma\) controls the strength of the repulsive force.

The optimization is performed using manual stochastic coordinate updates rather than automatic differentiation. At each epoch, the positive edges are randomly permuted and processed in mini-batches. This permutation only randomizes the order of positive-edge updates. For each positive edge \((i,j)\), the algorithm samples a fixed number of negative vertices uniformly from the dataset, controlled by the parameter \texttt{negative\_sample\_rate}. Thus, each positive update is coupled with several repulsive updates, following the negative-sampling strategy used in UMAP  graph-layout optimization~\cite{umap_2020,damrich2021umaploss} inspired by \cite{word2vec2013}.

\noindent For a positive edge \((i,j)\), the attractive update is proportional to
\[
-\frac{2ab\|y_i-y_j\|_2^{2b-2}}
        {1+a\|y_i-y_j\|_2^{2b}}
        (y_i-y_j),
\]
and is scaled by the normalized graph weight \(\widetilde p_{ij}\). For a negative sample \((i,k)\), the repulsive update is proportional to
\[
\frac{2\gamma b}
     {(\varepsilon+\|y_i-y_k\|_2^2)
      \left(1+a\|y_i-y_k\|_2^{2b}\right)}
      (y_i-y_k),
\]
where \(\varepsilon>0\) is a small numerical constant. The learning rate is linearly decreased during training, and gradients are clipped to improve numerical stability.

Although the optimizer is inspired by UMAP \cite{umap_2020}, there is one implementation difference. In the reference UMAP optimizer, edge weights affect the frequency with which positive edges are sampled \cite{umap_2020, UMAPDocsHowUMAPWorks}. But in CosMAP, all positive edges are visited once per epoch after random permutation, and the edge weight directly scales the attractive gradient. The negative sampling mechanism remains conceptually the same: random vertices are sampled as negative examples to approximate the repulsive part of the binary cross-entropy objective.

\subsection{Main Algorithm for CosMAP}
\begin{algorithm}[H]
\caption{CosMAP optimization by mini-batch  SGD with negative sampling}
\label{alg:optimization_cosmap}
\small
\DontPrintSemicolon

\KwIn{
edge endpoints $(h_k,t_k)$ from $P$ and weights $p_k=p_{h_k t_k}$, for $k \in \{1,\ldots,|E|\}$;
initial embedding $\mathbf{Y}\in\mathbb{R}^{n\times d}$;
number of epochs $T$; initial learning rate $\alpha_0$; batch size $B$;
negative sampling rate $m$; repulsion weight $\gamma$;
curve parameters $a,b$; gradient clipping value $c$;
distance floor $\epsilon_d=10^{-6}$; repulsive offset $\epsilon_r=10^{-3}$.
}

\KwOut{optimized embedding $\mathbf{Y}$.}

Normalize edge weights:

$\widetilde p_{k} \leftarrow  p_{k} / \max_j  p_{j}$ for all $j \in \{1,\ldots,|E|\}.$\;

\For{$t=0$ \KwTo $T-1$}{
    $\displaystyle
    \alpha_t \leftarrow
    \max\!\left(\alpha_0(1-t/T),\,10^{-3}\alpha_0\right)$\;

    Randomly permute the edge indices $\{1,\ldots,|E|\}$\;

    \ForEach{mini-batch $\mathcal{B}$ of size at most $B$}{
        Gather positive pairs $(h_k,t_k)$ and weights $\widetilde p_k$ for $k\in\mathcal{B}$\;

        $\mathbf{d}_k \leftarrow \mathbf{y}_{h_k}-\mathbf{y}_{t_k}$\;

        $\displaystyle
        d_{+,k}^2
        \leftarrow
        \max\!\left(\|\mathbf{d}_k\|^2,\epsilon_d\right)$\;

        $\displaystyle
        \phi_{+,k}
        \leftarrow
        \frac{-2ab(d_{+,k}^2)^{b-1}}
        {1+a(d_{+,k}^2)^b}$\;

        $\displaystyle
        \mathbf{g}_{+,k}
        \leftarrow
        \operatorname{clip}\!\left(
        \phi_{+,k}\mathbf{d}_k,-c,c
        \right)$\;

        $\displaystyle
        \mathbf{g}_{+,k}
        \leftarrow
        \widetilde p_k \odot\mathbf{g}_{+,k}$\;

        Accumulate attractive updates:
  \[
\left\{
\begin{aligned}
\mathbf{Y}[h_k] &\leftarrow \mathbf{Y}[h_k] + \alpha_t \mathbf{g}_{+,k}, \\
\mathbf{Y}[t_k] &\leftarrow \mathbf{Y}[t_k] - \alpha_t \mathbf{g}_{+,k},
\end{aligned}
\right. \quad k \in \mathcal{B}
\]

        \If{$m>0$}{
            Repeat each source vertex $h_k$ exactly $m$ times to obtain negative sources $\tilde h_j$\;

            Sample negative targets
            \[
            \tilde s_j \sim \operatorname{Uniform}(\{0,\ldots,n-1\}).
            \]

            \ForEach{$j$ such that $\tilde s_j=\tilde h_j$}{
                $\tilde s_j \leftarrow (\tilde s_j+1)\bmod n$\;
            }

            $\mathbf{d}'_j
            \leftarrow
            \mathbf{y}_{\tilde h_j}-\mathbf{y}_{\tilde s_j}$\;

            $\displaystyle
            d_{-,j}^2
            \leftarrow
            \max\!\left(\|\mathbf{d}'_j\|^2,\epsilon_d\right)$\;

            $\displaystyle
            \phi_{-,j}
            \leftarrow
            \frac{2\gamma b}
            {(\epsilon_r+d_{-,j}^2)
            \left(1+a(d_{-,j}^2)^b\right)}$\;

            $\displaystyle
            \mathbf{g}_{-,j}
            \leftarrow
            \operatorname{clip}\!\left(
            \phi_{-,j}\mathbf{d}'_j,-c,c
            \right)$\;

            Accumulate repulsive updates only on the negative source vertices:
            \[
            \mathbf{Y}[\tilde h_j] \leftarrow \mathbf{Y}[\tilde h_j]
            \mathrel{+}
            \alpha_t\mathbf{g}_{-,j}.
            \]
        }
    }
}

\KwRet{$\mathbf{Y}$}
\end{algorithm}

\newpage
\noindent The refinement phase in CosMAP first learns an intermediate representation in a dimension \(r>d\), and then uses a coordinate projection of this representation to initialize the final \(d\)-dimensional embedding. Now let's understand the first  phase of CosMAP, by showing this pseudo code below:

\begin{algorithm}[H]
\caption{CosMAP}
\label{alg:cosmap}
\small

\textbf{Input:} data matrix $\mathbf{X}$, number of neighbors $k$, embedding dimension $d$.

\textbf{Ensure:} low-dimensional embedding $\mathbf{Y}\in\mathbb{R}^{n\times d}$.

\begin{enumerate}
    \item Initialize $\mathbf{Y}$ using PCA, spectral embedding, random initialization,
    or a user-provided array.
    
    \item Construct the $k$-nearest-neighbor graph from $\mathbf{X}$ by using the user-provided metric.
    
    \item Compute the symmetric affinity weights $p_{ij}$ according to
    Equation~\ref{eq:pij_sym}.
    
    \item Optimize $\mathbf{Y}$ by minimizing the  sampled binary cross-entropy objective
    defined in Equation~\ref{eq:sampled_bce} using the Algorithm ~\ref{alg:optimization_cosmap}.
\end{enumerate}

\textbf{Return:} $\mathbf{Y}$.

\end{algorithm}

\vspace{0.6cm}

\subsection{CosMAP refinement pipeline Algorithm }

\begin{algorithm}[H]
\caption{CosMAP refinement pipeline}
\label{alg:cosmap_refinement}
\small

\textbf{Input:} data matrix $\mathbf{X}$, number of neighbors $k$, embedding dimension $d$, 
intermediate dimension $r>d$.

\textbf{Ensure:} final $d$-dimensional embedding
$\mathbf{Y}_{\mathrm{final}}\in\mathbb{R}^{n\times d}$.

\noindent \begin{enumerate}
    \item Apply Algorithm~\ref{alg:cosmap} to $\mathbf{X}$ with embedding
    dimension $r$, using PCA, spectral initialization, random initialization,
    or a user-provided array:
    \[
        \mathbf{X}
        \xrightarrow[\mathrm{CosMAP}]{\mathrm{dim}=r}
        \mathbf{Y}_{r}.
    \]

    \item Initialize the second phase with the first $d$ coordinates of
    $\mathbf{Y}_{r}$:
    \[
        \mathbf{Y}^{(d)} = \mathbf{Y}_{r}[:,1\!:\!d].
    \]

    \item Apply Algorithm~\ref{alg:cosmap} again, using $\mathbf{Y}_{r}$ as input
    and $\mathbf{Y}^{(d)}$ as initialization:
    \[
        \mathbf{Y}_{r}
        \xrightarrow[\mathrm{CosMAP}]{\mathrm{dim}=d,\; \mathrm{init}=\mathbf{Y}^{(d)}}
        \mathbf{Y}_{\mathrm{final}}.
    \]
\end{enumerate}

\textbf{Return:} $\mathbf{Y}_{\mathrm{final}}$.

\end{algorithm}

\vspace{0.5cm}
\noindent In practice, \(r\) is chosen larger than the final visualization dimension, for example \(r=30\) or \(r=50\), while \(d=2\) for planar visualization.

\newpage

\begin{center}
    \Huge\textbf{Appendix B}
\end{center}
\label{appendix_b}

\section{Discussion}\label{Discussion}

\subsection{Ablation study}

As discussed in the methodology section \ref{sec:method}, cosine similarity and Euclidean distance are closely related once vectors are $\ell_2$-normalized, see the equation \ref{eq:cos_vs_euc}. This suggests that, in principle, we could replace cosine in the high-dimensional similarity with a function of the Euclidean distance and keep the same NT-Xent-style temperature scaling \cite{simclr}. To test how crucial the metric  used in CosMAP is, we performed an ablation in which we removed cosine similarity and instead defined the high-dimensional conditional distribution using a Gaussian kernel on Euclidean distances:
\begin{equation}
    P_{j\mid i} =
    \begin{cases}
        \displaystyle
        \frac{\exp\!\bigl(-d(x_i, x_j)^2 / (2\tau)\bigr)}
             {\displaystyle\sum_{l\in\mathcal{N}_k(i)}\exp\!\bigl(-d(x_i, x_l)^2 / (2\tau)\bigr)},
        & j \in \mathcal{N}_k(i),\\[0.5em]
        0, & \text{otherwise},
    \end{cases}
    \label{eq:euclidean-ablation}
\end{equation}
where $d(x_i,x_j)$ is the Euclidean distance and $\tau$ plays the role of a fixed temperature, in contrast to the data-adaptive bandwidths used in t-SNE and related methods \cite{vandermaaten2008tsne,umap_2020}.

\begin{figure*}[h!]
    \centering
    \includegraphics[width=\textwidth]{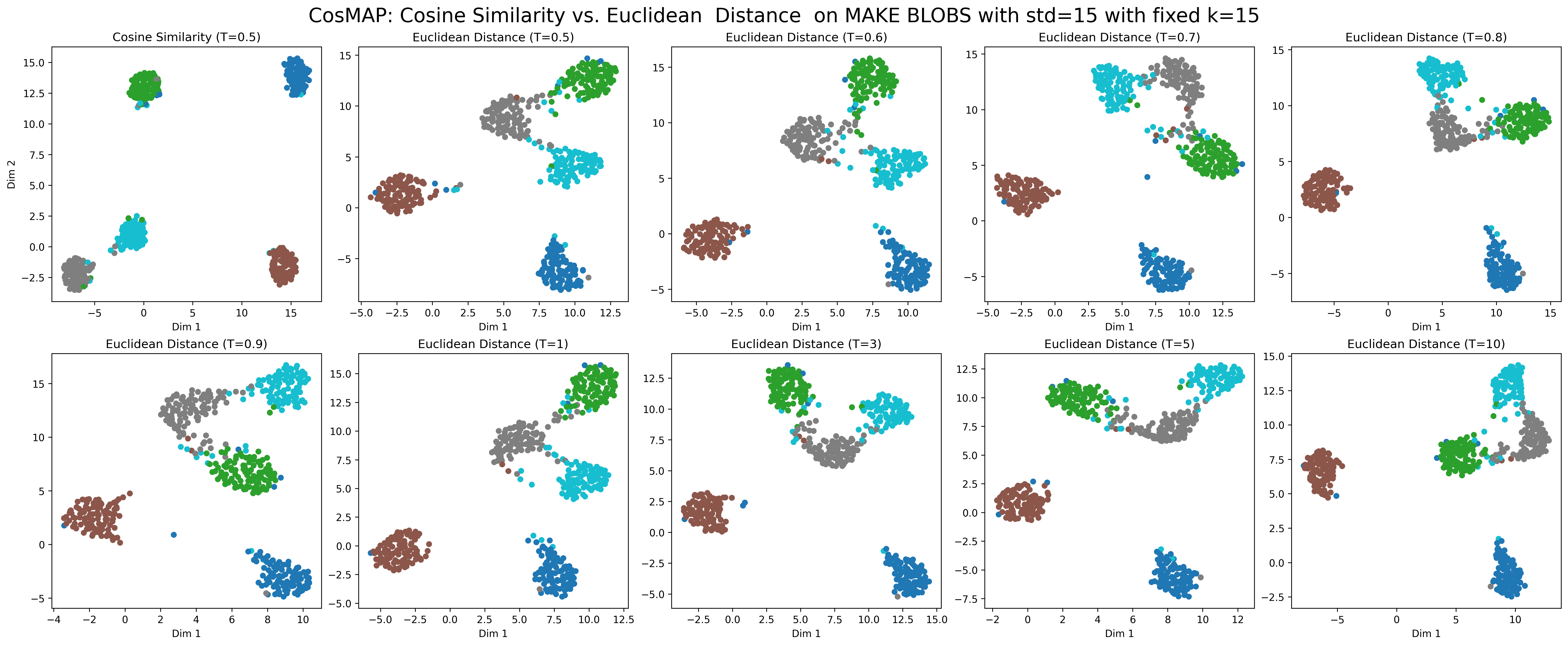}
    \caption{Sensitivity analysis on temperature parameters with Euclidean distance (\texttt{cluster\_std}=15) with a fixed number of neighbors ($k=15$).}
    \label{fig:blob_std15_temperature}
\end{figure*}

For this ablation, we used a controlled synthetic dataset generated with \texttt{make\_blobs} from \texttt{scikit-learn}, with $n = 500$ samples, $d = 100$ features, $5$ Gaussian clusters, and a cluster standard deviation of $15$ with the same mean. This synthetic setting was used deliberately because the goal was not to reproduce the complexity of real datasets such as MNIST,  single-cell RNA-seq data, or genealogical data, but rather to isolate the effect of the metric used in the high-dimensional similarity graph. In real datasets, several factors can interact with the embedding quality, including class overlap, biological heterogeneity, sparsity, batch effects, and non-Gaussian cluster shapes. By contrast, \texttt{make\_blobs} provides a simple setting in which the true cluster structure is known, allowing us to evaluate whether the behaviour of CosMAP is mainly driven by the choice of metric and hyperparameters.

Figure~\ref{fig:blob_std15_temperature} first compares CosMAP with cosine similarity at the default setting $(\tau=0.5,k=15)$ against CosMAP with Euclidean distance while varying the temperature $\tau$. The cosine-based configuration produces a stable embedding with well-separated clusters. In the Euclidean case, several clusters are also partially separated, showing that Euclidean distance does not completely fail on this dataset. However, increasing the temperature from $\tau=0.5$ to $T=10$ does not substantially change the overall geometry of the embedding. Across temperatures, the Euclidean layouts preserve a similar curved or chain-like organization, and some groups remain connected by intermediate points. In particular, the green, light green, and grey clusters are not consistently separated. This suggests that, for this dataset, the temperature parameter alone is not sufficient to correct the geometric distortions introduced by the Euclidean neighbourhood graph.

\begin{figure*}[h!]
    \centering
    \includegraphics[width=\textwidth]{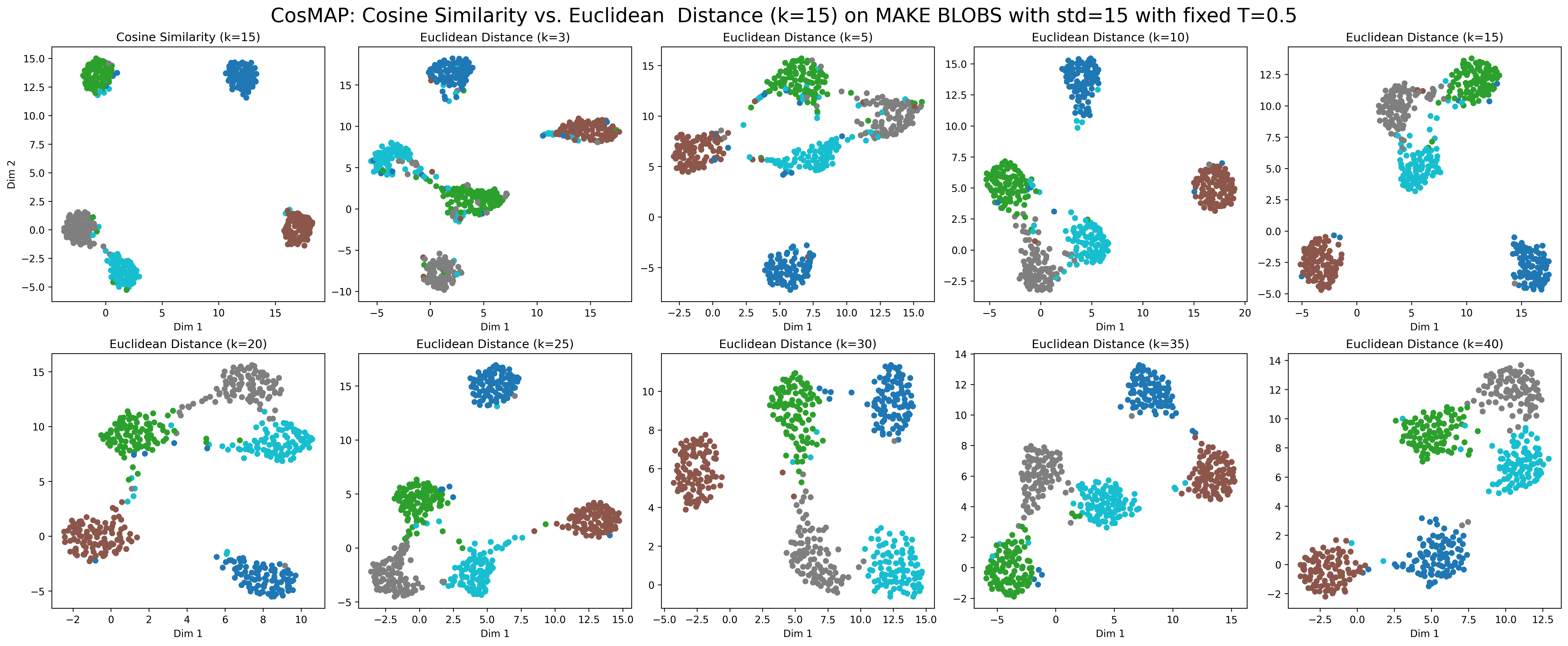}
    \caption{Sensitivity analysis on temperature parameters with Euclidean distance (\texttt{cluster\_std}=15) with a fixed temperatue ($\tau=0.5$).}
    \label{fig:blob_std15_fixed_temperature}
\end{figure*}

We therefore further investigated the influence of the neighbourhood size $k$ while keeping the temperature fixed at $\tau=0.5$ (Figure~\ref{fig:blob_std15_fixed_temperature}). This experiment shows that the Euclidean metric in CosMAP is much more sensitive to $k$ than to the temperature. For small values of $k$, such as $k=3$ or $k=5$, several clusters are connected through artificial bridges. For intermediate and larger values of $k$ the separation, it improves in some cases, but the global arrangement changes considerably from one value of $k$ to another. Even when the clusters become more distinguishable, the Euclidean embeddings remain less stable, and some clusters, especially the green, light green, and grey groups, are still not cleanly separated across all settings.

These observations support the choice of cosine similarity as the default metric in CosMAP. The ablation does not show that Euclidean distance is unusable; rather, it shows that Euclidean distance requires more careful tuning of the neighbourhood size and produces embeddings whose geometry changes noticeably across hyperparameter settings. In contrast, cosine similarity with the default configuration $(\tau=0.5,k=15)$ gives a more stable and visually coherent embedding.

Finally, regarding this contrastive temperature, we fix $\tau = 0.5$. This choice is motivated by the NT-Xent-style formulation used in contrastive learning, where the temperature controls the sharpness of the similarity distribution and the relative contribution of hard negatives \cite{simclr}. More importantly, this value is also confirmed by this ablation study. On the controlled \texttt{make\_blobs} dataset, CosMAP with cosine similarity and the default setting $\tau = 0.5$ produces a stable and well-separated embedding. By contrast, when Euclidean distance is used in the high-dimensional graph, varying the temperature over a wide range does not substantially correct the global geometry of the embedding: the layouts remain similar in shape, and some clusters remain connected or only partially separated. This suggests that, in CosMAP, the temperature is not the main source of instability. This ablation therefore supports $\tau = 0.5$ as a reasonable default temperature for any metric used in  CosMAP.

\subsection{Sensitivity analysis}
\begin{figure*}[h!]
    \centering
    \includegraphics[width=\textwidth]{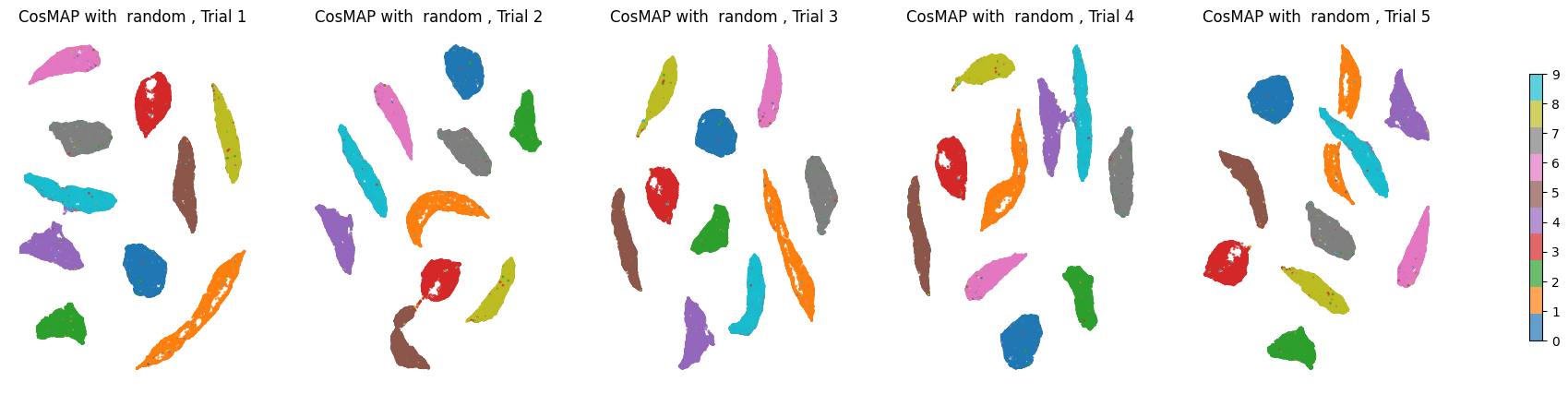}
    \caption{CosMAP under random initialisation with 5 different trials on MNIST dataset}
    \label{fig:cosmap_random_trials}
\end{figure*}

Contrastive Neighbour-embedding methods are generally sensitive to initialization and hyperparameter choices, since the final two-dimensional layout is obtained by optimizing a non-convex objective on a neighbourhood graph \cite{umap_2020,wang2021pacmap}. In CosMAP, initialization affects the stability of the final embedding because the optimization is stochastic and relies on negative sampling. In preliminary experiments, PCA initialization often produced fragmented layouts, especially on large and heterogeneous single-cell datasets. We therefore use a two-dimensional spectral initialization computed from the graph Laplacian of the high-dimensional $k$-NN graph, following the strategy used in UMAP \cite{umap_2020}. This choice is natural for CosMAP because the initialization is derived from the same neighbourhood graph that is later optimized, whereas PCA only preserves directions of maximum variance in the original feature space.
Figure~\ref{fig:cosmap_random_trials} uses MNIST instead of the synthetic \texttt{make\_blobs} dataset because the two experiments address different questions. The \texttt{make\_blobs} dataset was used in the previous ablation to isolate the effect of the metric, temperature, and neighbourhood size under a controlled cluster structure. Here, the objective is to evaluate the sensitivity of CosMAP to random initialization on a real benchmark with  visually interpretable classes. Therefore, we run CosMAP on MNIST with random initialization and a different seed for each of the five trials. Although the resulting embeddings remain meaningful across runs, their global organization varies across trials. Some clusters are not consistently separated, while others are split into multiple subclusters depending on the random seed. This confirms that the initial configuration can influence the final layout.
 . For this reason, all comparative experiments use spectral initialization by default. This contrasts with methods such as PaCMAP and LocalMAP, which explicitly control the contribution of neighbour, mid-near, and further pairs during training to balance local and global structure using random initialization \cite{wang2021pacmap, wang2025localmap}.

For the low-dimensional similarity kernel, we use the same calibration strategy as UMAP to determine the shape parameters $(a,b)$ from the chosen \texttt{min\_dist} and \texttt{spread} values \cite{umap_2020}. These parameters control the effective compactness of clusters and the heaviness of the tail in the low-dimensional similarity $q_{ij}$. In all experiments, we keep these parameters fixed in order to avoid introducing additional sources of variation and to focus the analysis on the effect of the high-dimensional similarity graph.

The neighbourhood size $k$ controls the scale at which the high-dimensional graph is constructed, and therefore determines the balance between local and more global structure in the final embedding. We use $k=15$ as the default value, which is consistent with common neighbourhood-embedding practice and provides a reasonable compromise between an overly sparse graph and an overly smoothed graph \cite{umap_2020}. Very small values of $k$ may fragment the neighbourhood graph, whereas very large values may connect points that are not truly close in the intrinsic geometry of the data, especially in high-dimensional settings where distance concentration can make Euclidean neighbourhoods less informative \cite{aggarwal2001surprising,watson_how_2022}. In our case, the ablation study confirms that the default configuration $(\tau=0.5,k=15)$ is appropriate for the cosine similarity  in CosMAP. Therefore, in the our experiments results above, we kept these parameters fixed in order to evaluate the behaviour of the method across datasets without introducing dataset-specific hyperparameter tuning.

\newpage

\begin{center}
    \Huge\textbf{Appendix C}
\end{center}

\section{Additional results for CosMAP against the state-of-the-art}

\label{appendix_c}

\subsection{Heart Cell Atlas dataset}
\label{sec:heart_cell_atlas_dataset}

The Heart Cell Atlas dataset contains single-cell and single-nucleus RNA-seq
profiles from human cardiac tissue across multiple anatomical regions
\cite{litvinukova2020heart}. In this benchmark, we use the subsampled version
distributed through the \texttt{scvi-tools} dataset API
\cite{gayoso_scvi_tools_2021}. This dataset of  size $18641 \times 1200$ represents a larger and more
heterogeneous organ-scale benchmark than the Cortex dataset, making it useful for
evaluating the behavior of dimensionality-reduction methods on complex biological
structure. 
\begin{figure*}[h!]
	\centering
	\includegraphics[width=\textwidth]{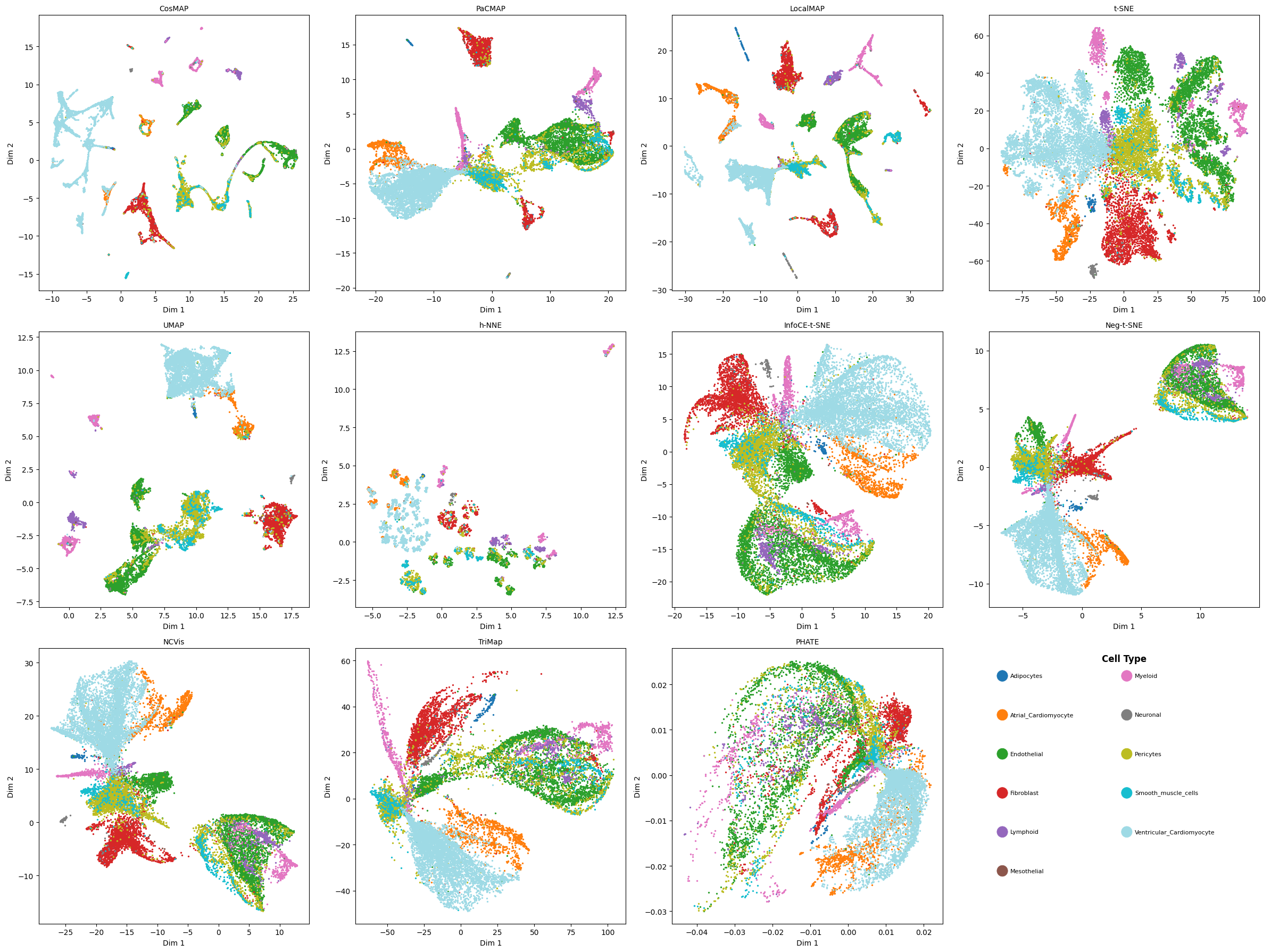}
	\caption{Comparison of DR methods on the Heart Cell Atlas dataset.}
	\label{fig:cosmap-vs-sota-heart-cell-atlas}
\end{figure*}

\newpage

\subsection{Peripheral blood mononuclear cells dataset}
\label{sec:pbmc_dataset}

The peripheral blood mononuclear cells (PBMC) dataset is commonly used as a benchmark for single-cell RNA-seq analysis. It provides a heterogeneous
immune-cell setting in which dimensionality-reduction methods are expected to separate major immune populations while preserving related cell-type relationships. 

\begin{figure*}[h!]
	\centering
	\includegraphics[width=\textwidth]{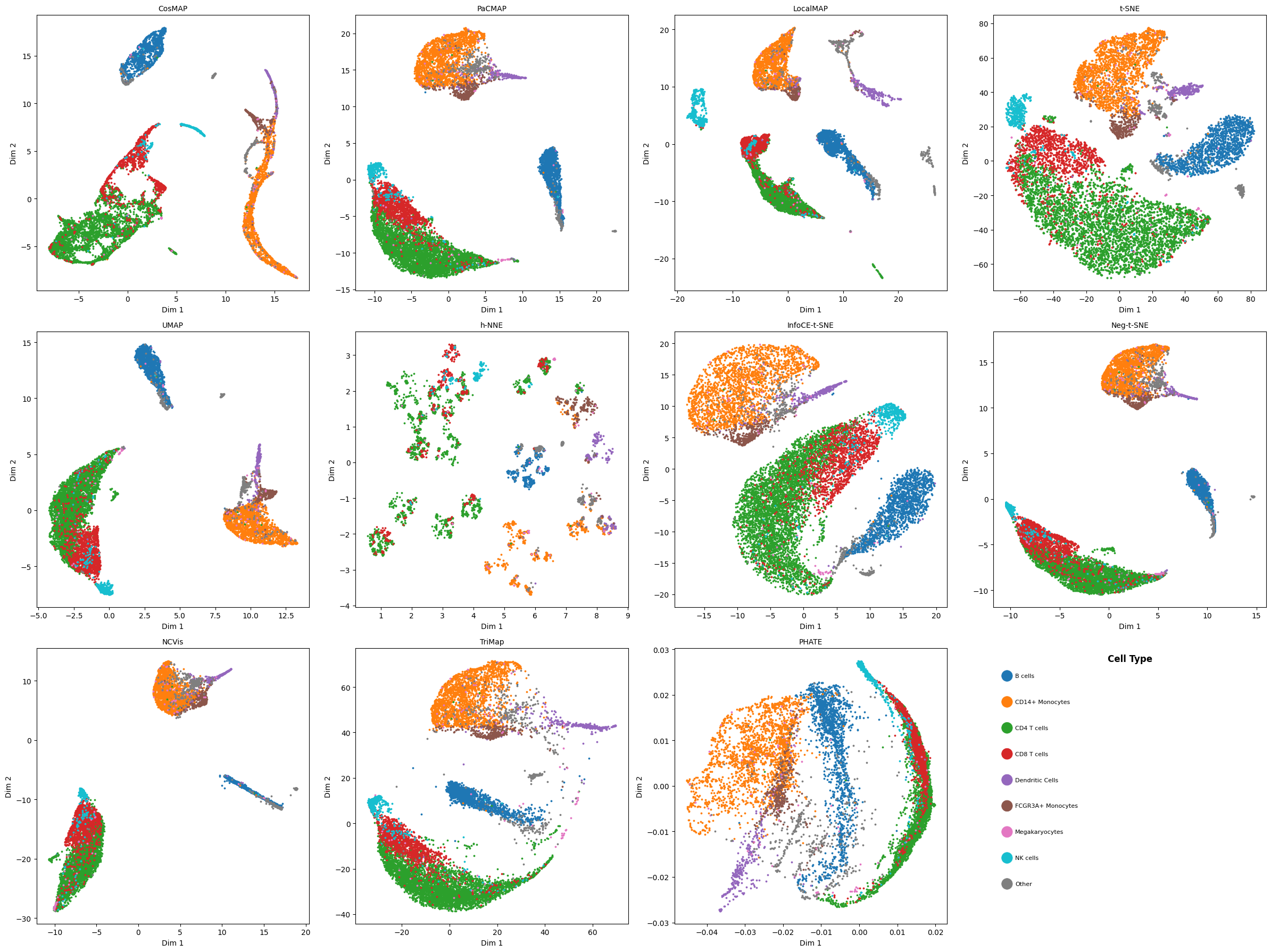}
	\caption{Comparison of DR methods on the peripheral blood mononuclear cells dataset.}
	\label{fig:cosmap-vs-sota-pbmc}
\end{figure*}
\newpage

\subsection{COIL-20 dataset}
\label{sec:coil20_dataset}

\begin{figure*}[h!]
	\centering
	\includegraphics[width=\textwidth]{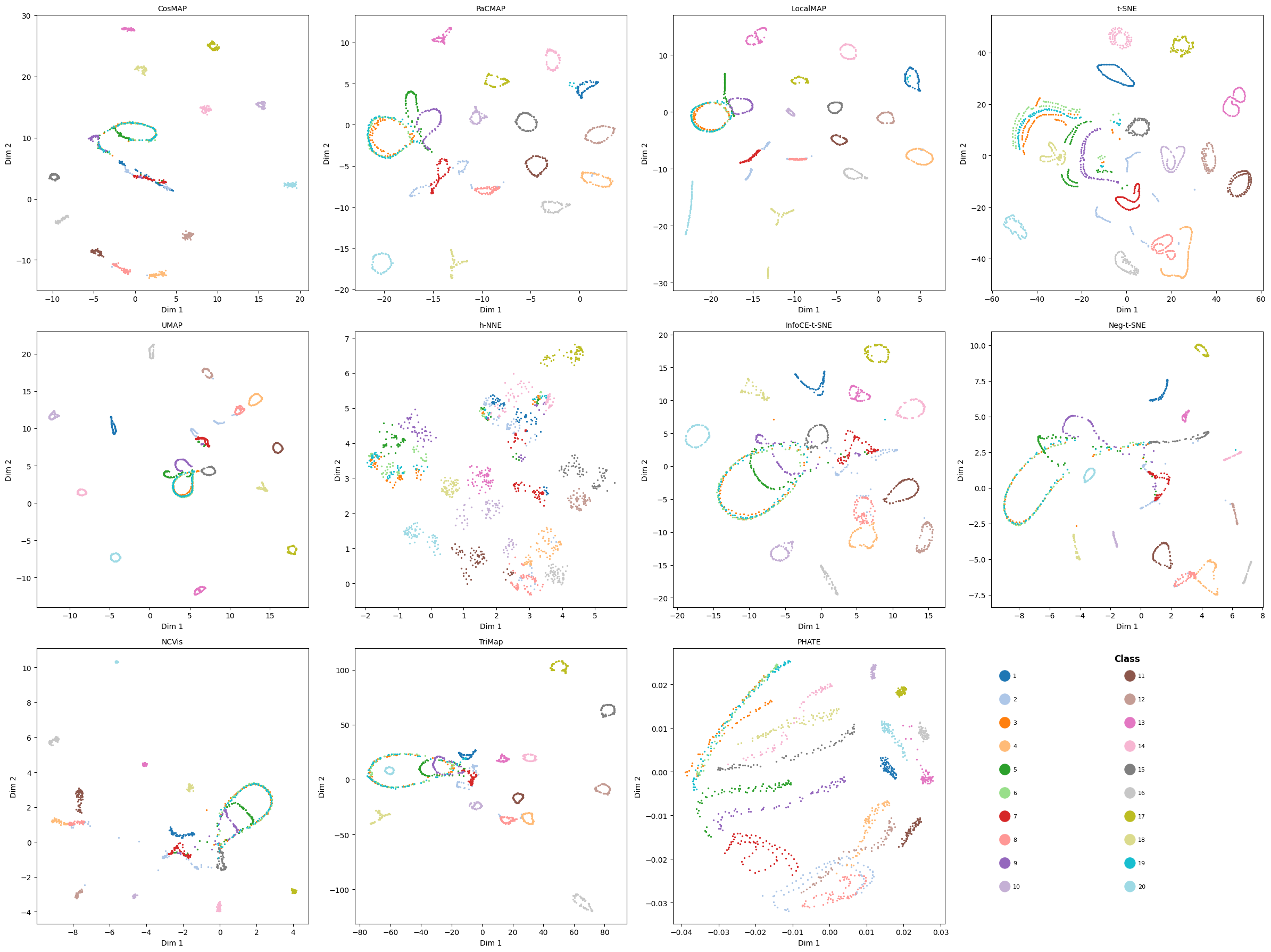}
	\caption{Comparison of DR methods on the COIL-20 dataset.}
	\label{fig:cosmap-vs-sota-coil-20}
\end{figure*}

 In this image ~\ref{fig:cosmap-vs-sota-coil-20}, we use the preprocessed COIL-20 data provided in the \href{https://github.com/williamsyy/LocalMAP/tree/experiments/data}{LocalMAP GitHub repository} , where the flatted images and labels are loaded from NumPy files, \texttt{coil\_20.npy} and \texttt{coil\_20\_labels.npy}. A good embedding should therefore form well-separated object clusters, while ideally preserving some smooth within-class structure induced by the rotation angle.

\end{document}